\documentclass[preprint2]{aastex631}
\usepackage{CJK}
\usepackage{textcomp}
\DeclareTextSymbolDefault{\dh}{T1}

\shortauthors{Ca\~nas et al.}
\shorttitle{Two Gas Giants transiting M dwarfs}
\graphicspath{{./}{figures/}}

\newcommand{\UA}{Steward Observatory, The University of Arizona, 933 N.\ Cherry Avenue, Tucson, AZ 85721, USA}
\newcommand{\UAUG}{Department of Astronomy, The University of Arizona, 933 N.\ Cherry Avenue, Tucson, AZ 85721, USA}
\newcommand{\PSUAA}{Department of Astronomy \& Astrophysics, The Pennsylvania State University, 525 Davey Laboratory, University Park, PA 16802, USA}
\newcommand{\PSUCEHW}{Center for Exoplanets and Habitable Worlds, The Pennsylvania State University, 525 Davey Laboratory, University Park, PA 16802, USA}
\newcommand{\Princeton}{Department of Astrophysical Sciences, Princeton University, 4 Ivy Lane, Princeton, NJ 08540, USA}
\newcommand{\RUSSELL}{Henry Norris Russell Fellow} 
\newcommand{\Macquarie}{Department of Physics and Astronomy, Macquarie University, Balaclava Road, North Ryde, NSW 2109, Australia}
\newcommand{\UCI}{Department of Physics \& Astronomy, The University of California, Irvine, Irvine, CA 92697, USA}
\newcommand{\Carleton}{Department of Physics and Astronomy, Carleton College, One North College Street, Northfield, MN 55057, USA}
\newcommand{\NIST}{Time and Frequency Division, National Institute of Standards and Technology, 325 Broadway, Boulder, CO 80305, USA}
\newcommand{\NISTAssoc}{Associate of the National Institute of Standards and Technology, 325 Broadway, Boulder, CO 80305, USA}
\newcommand{\CUBoulder}{Department of Physics, University of Colorado, 2000 Colorado Avenue, Boulder, CO 80309, USA}
\newcommand{\UT}{McDonald Observatory and Center for Planetary Systems Habitability, The University of Texas at Austin, Austin, TX 78730, USA}

\newcommand{\UWY}{Department of Physics \& Astronomy, University of Wyoming, Laramie, WY 82070, USA}
\newcommand{\STSCI}{Space Telescope Science Institute, 3700 San Martin Drive, Baltimore, MD 21218, USA}
\newcommand{\JH}{Department of Physics and Astronomy, Johns Hopkins University, 3400 N Charles St, Baltimore, MD 21218, USA}
\newcommand{\AFRL}{Space Vehicles Directorate, Air Force Research Laboratory, 3550 Aberdeen Ave. SE, Kirtland AFB, NM 87117, USA}
\newcommand{\NOIRLab}{NSF's National Optical-Infrared Astronomy Research Laboratory, 950 N. Cherry Ave., Tucson, AZ 85719, USA}
\newcommand{\OU}{Homer L. Dodge Department of Physics and Astronomy, University of Oklahoma, 440 W. Brooks Street, Norman, OK 73019, USA}
\newcommand{\TIFR}{Department of Astronomy and Astrophysics, Tata Institute of Fundamental Research, Homi Bhabha Road, Colaba, Mumbai 400005, India}
\newcommand{\HWS}{Department of Physics, Hobart and William Smith Colleges, 300 Pulteney Street, Geneva, NY 14456, USA}
\newcommand{\UPenn}{Department of Physics and Astronomy, University of Pennsylvania, 209 South 33rd Street, Philadelphia, PA 19104}
\newcommand{\PSETI}{Penn State Extraterrestrial Intelligence Center, 525 Davey Laboratory, The Pennsylvania State University, University Park, PA, 16802, USA}

\begin{document}
\begin{CJK*}{UTF8}{gbsn}

\title{TOI-3714 b and TOI-3629 b: Two gas giants transiting M dwarfs confirmed with HPF and NEID}
\correspondingauthor{Caleb I. Ca\~nas}
\email{canas@psu.edu}

\author[0000-0003-4835-0619]{Caleb I. Ca\~nas}
\altaffiliation{NASA Earth and Space Science Fellow}
\affiliation{\PSUAA}
\affiliation{\PSUCEHW}

\author[0000-0001-8401-4300]{Shubham Kanodia}
\affiliation{\PSUAA}
\affiliation{\PSUCEHW}

\author[0000-0003-4384-7220]{Chad F.\ Bender}
\affil{\UA}

\author[0000-0001-9596-7983]{Suvrath Mahadevan}
\affil{\PSUAA}
\affil{\PSUCEHW}

\author[0000-0001-7409-5688]{Gu\dh mundur Stef\'ansson}
\altaffiliation{\RUSSELL}
\affil{\Princeton}

\author[0000-0001-9662-3496]{William D. Cochran}
\affil{\UT}

\author[0000-0002-9082-6337]{Andrea S.J. Lin}
\affil{\PSUAA}
\affil{\PSUCEHW}

\author[0000-0003-4250-4437]{Hsiang-Chih Hwang}
\affiliation{School of Natural Sciences, Institute for Advanced Study, Princeton, 1 Einstein Drive, NJ 08540, USA}

\author[0000-0002-5300-5353]{Luke Powers}
\affil{\PSUAA}
\affil{\PSUCEHW}

\author[0000-0002-0048-2586]{Andrew Monson}
\affil{\PSUAA}

\author[0000-0003-3688-2298]{Elizabeth M. Green}
\affil{\UA}

\author[0000-0001-9307-8170]{Brock A. Parker}
\affil{\UWY}

\author[0000-0002-5817-202X]{Tera N. Swaby}
\affil{\UWY}

\author[0000-0002-4475-4176]{Henry A. Kobulnicky}
\affil{\UWY}

\author[0000-0001-9209-1808]{John Wisniewski}
\affiliation{\OU}

\author[0000-0002-5463-9980]{Arvind F. Gupta}
\affil{\PSUAA}
\affil{\PSUCEHW}

\author[0000-0002-0885-7215]{Mark E. Everett}
\affil{\NOIRLab}

\author[0000-0002-7227-2334]{Sinclaire Jones}
\affiliation{\Princeton}

\author[0000-0001-7614-9205]{Benjamin Anjakos}
\affil{\UAUG}

\author[0000-0001-7708-2364]{Corey Beard}
\affil{\UCI}

\author[0000-0002-6096-1749]{Cullen H.\ Blake}
\affil{\UPenn}

\author[0000-0002-2144-0764]{Scott A. Diddams}
\affil{\NIST}
\affil{\CUBoulder}

\author[0000-0001-9774-7802]{Zehao Dong (董泽浩)}
\affil{\UAUG}

\author[0000-0002-0560-1433]{Connor Fredrick}
\affil{\NISTAssoc}
\affil{\CUBoulder}

\author[0000-0003-0112-9095]{Elnaz Hakemiamjad}
\affil{\UAUG}

\author[0000-0003-1263-8637]{Leslie Hebb}
\affiliation{\HWS}

\author[0000-0002-2990-7613]{Jessica E. Libby-Roberts}
\affil{\PSUAA}
\affil{\PSUCEHW}

\author[0000-0002-9632-9382]{Sarah E. Logsdon}
\affil{\NOIRLab}

\author[0000-0003-0241-8956]{Michael W. McElwain}
\affil{Exoplanets and Stellar Astrophysics Laboratory, NASA Goddard Space Flight Center, Greenbelt, MD 20771, USA}

\author[0000-0001-5000-1018]{Andrew J. Metcalf}
\affil{\AFRL}
\affil{\NIST}
\affil{\CUBoulder}

\author[0000-0001-8720-5612]{Joe P.\ Ninan}
\affil{\TIFR}

\author[0000-0002-2488-7123]{Jayadev Rajagopal}
\affil{\NOIRLab}

\author[0000-0002-4289-7958]{Lawrence W. Ramsey}
\affil{\PSUAA}
\affil{\PSUCEHW}

\author[0000-0003-0149-9678]{Paul Robertson}
\affil{\UCI}

\author[0000-0001-8127-5775]{Arpita Roy}
\affil{\STSCI}
\affil{\JH}

\author[0000-0003-1082-9720]{Jacob Ruhle}
\affil{\UAUG}

\author[0000-0002-4046-987X]{Christian Schwab}
\affil{\Macquarie}

\author[0000-0002-4788-8858]{Ryan C. Terrien}
\affil{\Carleton}

\author[0000-0001-6160-5888]{Jason T. Wright}
\affil{\PSUAA}
\affil{\PSUCEHW}
\affil{\PSETI}



\begin{abstract}

We confirm the planetary nature of two gas giants discovered by TESS to transit M dwarfs. TOI-3714 ($V=15.24,~J=11.74$) is an M2 dwarf hosting a hot Jupiter ($M_p=0.70 \pm 0.03~\mathrm{M_J}$ and $R_p=1.01 \pm 0.03~\mathrm{R_J}$) on an orbital period of $2.154849 \pm 0.000001$ days with a resolved white dwarf companion. TOI-3629 ($V=14.63,~J=11.42$) is an M1 dwarf hosting a hot Jupiter ($M_p=0.26 \pm 0.02~\mathrm{M_J}$ and $R_p=0.74 \pm 0.02~\mathrm{R_J}$) on an orbital period of $3.936551_{-0.000006}^{+0.000005}$ days. We characterize each transiting companion using a combination of ground-based and space-based photometry, speckle imaging, and high-precision velocimetry from the Habitable-zone Planet Finder and the NEID spectrographs. With the discovery of these two systems, there are now nine M dwarfs known to host transiting hot Jupiters. Among this population, TOI-3714 b ($T_{eq}=750\pm20$ K and $\mathrm{TSM}=98\pm7$) and TOI-3629 b ($T_{eq}=690\pm20$ K and $\mathrm{TSM}=80\pm9$) are warm gas giants amenable to additional characterization with transmission spectroscopy to probe atmospheric chemistry and, for TOI-3714, obliquity measurements to probe formation scenarios.

\end{abstract}



\section{Introduction} \label{sec:intro}
Short-period ($P<10$ days) Jupiter-sized ($R_{p}\ge8~\mathrm{R_{\oplus}}$) exoplanets, or hot Jupiters, are rare in the Galaxy. Results from radial velocity (RV) surveys \citep[e.g.,][]{Cumming2008,Mayor2011,Wright2012}, ground-based photometry surveys \citep[e.g.,][]{Obermeier2016}, and space-based surveys \citep{Howard2012,Petigura2018,Zhou2019a} have determined the occurrence rate for hot Jupiters orbiting Sun-like (FGK) dwarfs to be \(\lesssim 1\%\). Despite that over 400 hot Jupiters have been detected orbiting Sun-like stars, there is no consensus as to the origin mechanisms required to create this population of exoplanets \citep[see][]{Dawson2018}. Many hypotheses have been proposed to explain the origin of these planets, including star-planet interactions \citep[e.g.,][]{Wu2003,Petrovich2015a}, planet-planet interactions \citep[e.g.,][]{Naoz2011}, migration due to planet-disk interactions \citep[e.g.,][]{Lin1996}, high-eccentricity migration \citep[e.g.,][]{Rasio1996,Weidenschilling1996,Ford2008,Petrovich2015}, and \textit{in-situ} formation \citep[e.g.,][]{Boley2016,Batygin2016}. 

From analysis of the Kepler field \citep[e.g.,][]{Dressing2015,Mulders2015,Hardegree-Ullman2019,Hsu2020}, the occurrence rate of small ($1\mathrm{~R_\oplus}<R_p<4\mathrm{~R_\oplus}$) planets on short-period ($P<200$ days) orbits is larger for M dwarfs, the most abundant type of star in the Galaxy \citep[][]{Henry2018}, compared to Sun-like stars. The occurrence rate of these small planets also increases for later type M dwarfs. RV surveys have similarly revealed the abundance of low-mass planets ($1\mathrm{~M_\oplus}<M_p<10\mathrm{~M_\oplus}$) on short-period orbits ($P<200$ days) as companions to M dwarfs \citep[e.g.,][]{Bonfils2013,Tuomi2014,Tuomi2019,Sabotta2021}. Jupiter-like planets, however, are expected to be rare companions to M dwarfs under the theory of core accretion \citep[e.g.,][]{Laughlin2004,Ida2005,Kennedy2008}. In the core accretion model, a gas giant planet forms from a runaway process resulting in the rapid accretion of gas onto a planetary core \citep[e.g.,][]{Pollack1996,Ida2004,Hubickyj2005}. This model predicts a small number of gas giants orbiting M dwarfs, because the low surface density of an M dwarf protoplanetary disk would impede the formation of massive cores required for the onset of runaway gas accretion. 

To date, M dwarf RV surveys \citep[e.g.,][]{Endl2006,Bonfils2013,Tuomi2019,Sabotta2021} and photometric surveys \citep{Kovacs2013,Obermeier2016} have only been able to constrain the occurrence rate to $\lesssim1-2\%$ for hot Jupiters orbiting M dwarfs. Prior to this paper, there were seven hot Jupiters known to transit M dwarfs: Kepler-45 b \citep{Johnson2012}, HATS-6 b \citep{Hartman2015}, NGTS-1 b \citep{Bayliss2018}, HATS-71 b \citep{Bakos2020}, HATS-74A b and HATS-75b \citep{Jordan2022}, and TOI-3757 b \citep{Kanodia2022}. 

In this paper, we confirm the planetary nature of two gas giants transiting the M dwarfs TOI-3714 ($V=14.63$, $J=11.42$, $T=12.79$) and TOI-3629 ($V=15.24$, $J=11.74$, $T=13.18$) . We characterize each system using space and ground-based photometry, speckle imaging, and precision RVs obtained with the Habitable-zone Planet Finder \citep[HPF;][]{Mahadevan2012,Mahadevan2014} and NEID \citep[][]{Schwab2016,Halverson2016} spectrographs. We derive stellar parameters for the host stars using our HPF spectra and use the RVs measured from both HPF and NEID to confirm that each transiting companion is a hot Jupiter.

This paper is structured as follows: Section \ref{sec:observations} presents the photometric, imaging, and spectroscopic observations used to characterize each system. The characterization of the host stars and the best estimates of the stellar parameters are described in Section \ref{sec:stellarpar}. The modeling and analysis of the photometry and RVs are presented in Section \ref{sec:modelfit}. Section \ref{sec:discussion} provides further discussion of the nature of these planets and the feasibility for future study. We end with a summary of our key results in Section \ref{sec:summary}.

\section{Observations} \label{sec:observations}
\subsection{TESS}
TESS \citep{Ricker2015} observed TOI-3629 (TIC 455784423, Gaia EDR3 2881820324294985856) and TOI-3714 (TIC 155867025, Gaia EDR3 178924390478792320) in long-cadence mode (30-min cadence). TOI-3629 was observed during Sector 17 (2019 October 7 through 2019 November 2) and TOI-3714 was observed during Sector 19 (2019 November 27 through 2019 December 24). Similar to TOI-1899 \citep{Canas2020}, we identified TIC-455784423.01 as a planetary candidate using a custom pipeline to search for transiting candidates in short and long-cadence TESS data orbiting M dwarfs that were amenable to RV observations with HPF. At the time we searched TESS data, the ``quick-look pipeline'' (QLP) developed by \cite{Huang2020,Huang2020b} was releasing candidates from the southern TESS sectors. Our search was not designed for completeness but to identify a few ($\lesssim10$) M dwarfs with Jupiter-sized transiting companions that were most likely planetary in nature.

Briefly, this pipeline was developed to identify transiting companions to bright (TESS magnitude of $T<13$) M dwarfs ($T_{e}<4000$ K) from the catalog of cool dwarfs \citep[a value of \texttt{splists = cooldwarfs\_v8};][]{Muirhead2018} in the TESS input catalog \citep[TIC;][]{Stassun2019} that are observable from the Hobby-Eberly Telescope \citep[HET;][]{Ramsey1994,Ramsey1998} at McDonald Observatory ($-11^\circ<\delta<71^\circ$). These constraints resulted in an average of $\sim2000$ stars to process per sector. Our pipeline uses the \texttt{lightkurve} package \citep{LightkurveCollaboration2018} to detrend (i) short-cadence light curves provided by the TESS science processing operations center \citep[][]{Jenkins2016} and (ii) long-cadence derived from calibrated full-frame images using \texttt{eleanor} \citep{Feinstein2019} with a Savitzky-Golay filter. The pipeline searches for transit-like events in the detrended photometry using the box least-squares algorithm \citep{Kovacs2002} and models the transit signal following the formalism from \cite{Mandel2002} as implemented in the \texttt{batman} package \citep{Kreidberg2015}. The transit-like events are vetted for centroid offsets and inconsistencies ($>3\sigma$ discrepant) with the stellar density recovered by the transit fit \citep[e.g.,][]{Seager2003,Winn2010} and the stellar density reported by TIC. Signals that were identified were subsequently vetted by members of the HPF team before we began RV observations.

We detected one planet candidate with a depth of $\sim1.5\%$ and a period of $\sim3.94$ days. This event was subsequently identified (at a comparable period and depth) by the QLP and given the designation TOI-3629.01. It is one of the planetary candidates from the ``faint star search''\footnote{\url{https://tess.mit.edu/qlp/}}, an effort to extend the nominal search and vetting of TESS objects of interests to stars with a TESS magnitude of $T<13.5$ \citep{Kunimoto2021}. The faint star search also identified TOI-3714.01 as a transiting candidate with  a depth of $\sim4.5\%$ and a period of $\sim2.15$ days. This target was excluded from our search due to its faintness ($T=13.18$).

We extract the photometry from the TESS full-frame images using \texttt{eleanor}, which calls the TESScut\footnote{\url{https://mast.stsci.edu/tesscut/}} service \citep{Brasseur2019} to obtain a cut-out of \(31\times31\) pixels of the calibrated full-frame images centered on each target. \texttt{eleanor} removes the background, corrects for systematics, and derives a light curve for various combinations of apertures when processing a target. The final light curve is the one which minimizes the combined differential photometric precision (CDPP) after the data are binned to 1 hour timescales. The CDPP was originally defined for Kepler as the rms of the photometric noise on transit timescales \citep{Jenkins2010}. Minimizing this value ensures that sharp features on relatively short timescales, such as transits, are preserved. The final CDPP was 2902 ppm for TOI-3714 and 2219 ppm for TOI-3629.

\begin{figure*}[!ht]
\epsscale{1.15}
\plotone{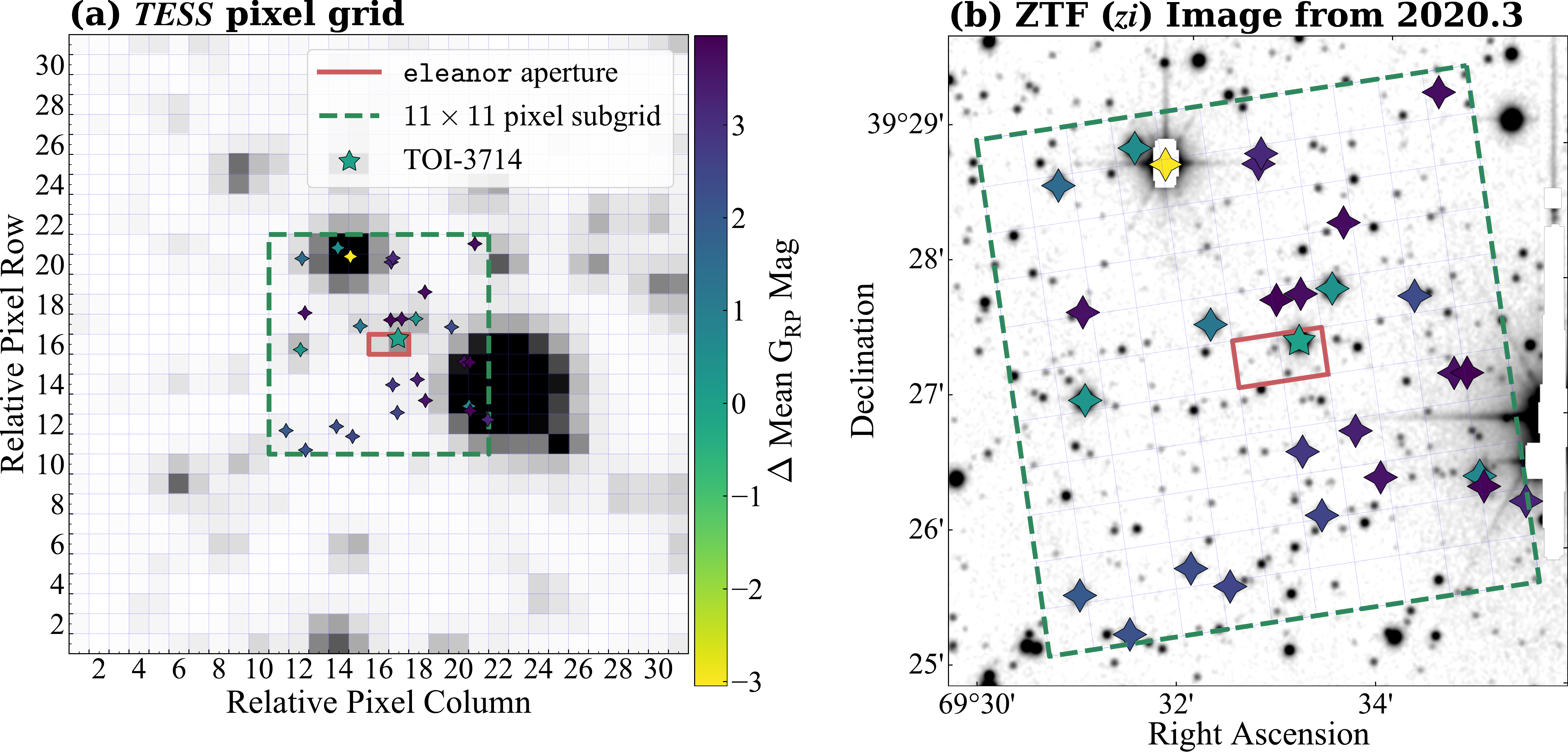}
\caption{\textbf{(a)} The $31\times31$ TESS target pixel cutout centered around TOI-3714 (marked as a star). Stars identified in Gaia EDR3 with magnitudes $\Delta\mathrm{G_{RP}}<4$ are marked with diamond stars for reference. Stars with $\Delta\mathrm{G_{RP}}<0$ are brighter than the host star. The dashed line denotes the TESS $11\times11$ pixel subgrid that is shown in (b). \textbf{(b)} Overlay of the TESS $11\times11$ pixel subgrid, TOI-3714, and other comparably bright stars on a ZTF $zi$ image.}
\label{fig:3714apertures}
\end{figure*}

\begin{figure*}[!ht]
\epsscale{1.15}
\plotone{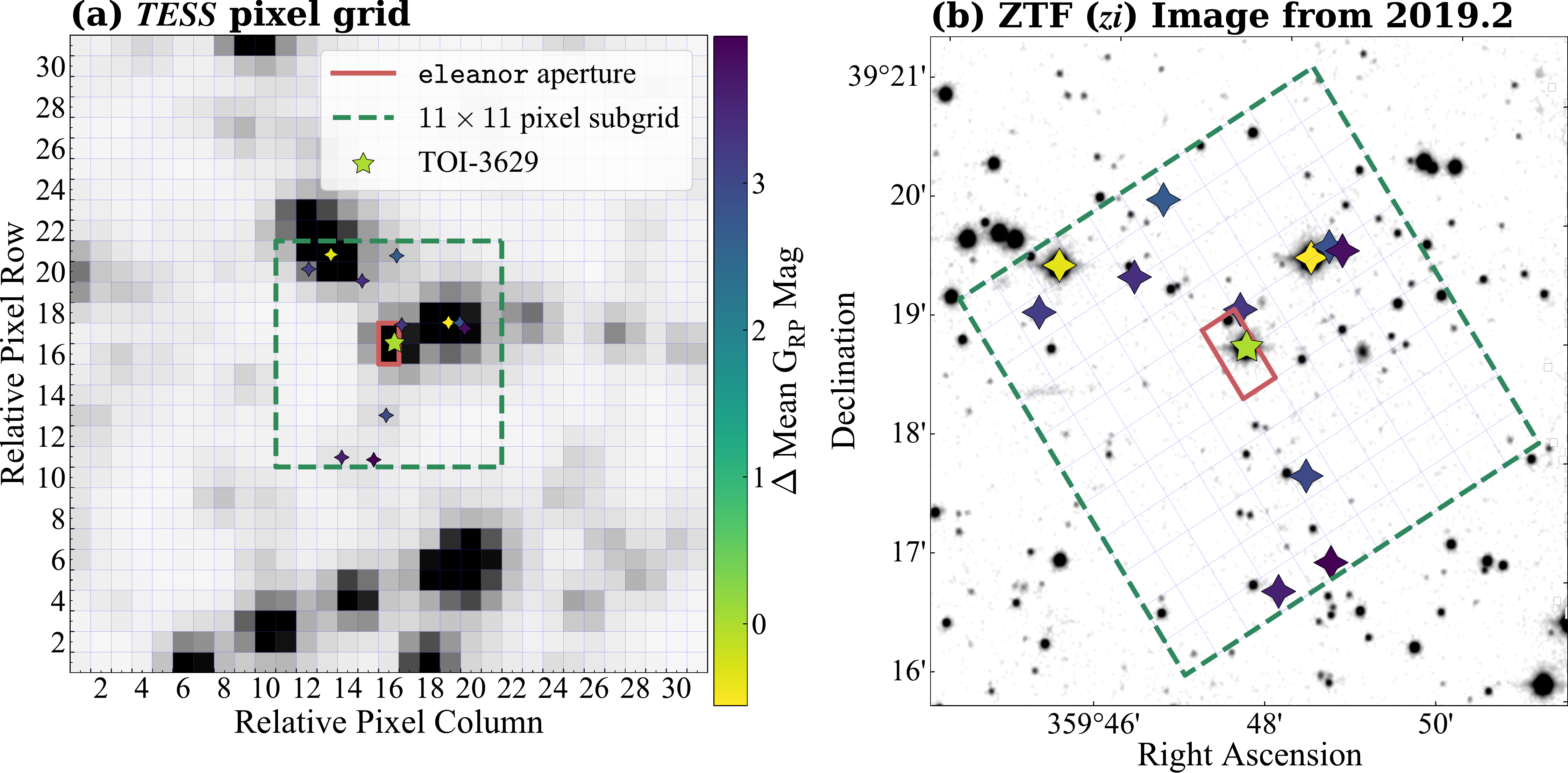}
\caption{Identical to Figure \ref{fig:3714apertures} but for TOI-3629. \textbf{(a)} The $31\times31$ TESS target pixel cutout centered around TOI-3629 (marked as a star). Stars identified in Gaia EDR3 having magnitudes $\Delta\mathrm{G_{RP}}<4$ are marked with diamond stars. Stars with $\Delta\mathrm{G_{RP}}<0$ are brighter than the host star. \textbf{(b)} Overlay of the TESS $11\times11$ pixel subgrid, TOI-3629, and other comparably bright stars on a ZTF $zi$ image.}
\label{fig:3629apertures}
\end{figure*}

Figures \ref{fig:3714apertures} and \ref{fig:3629apertures} show the photometric images for TOI-3629 and TOI-3714, respectively. Panel (a) in both figures presents the TESS full frame image cutouts and the apertures used to derive the light curves for each target. In panel (b), a smaller 11x11 pixel subgrid of the TESS image and the light curve apertures are overplotted on images from the  Zwicky Transient Facility \citep[ZTF;][]{Masci2019}. For each target, the preferred aperture is a $2\times1$ rectangular aperture centered on the host star. To investigate the impact of background stars as a source of dilution, we searched the $11\times11$ TESS pixel grid centered on each target in Gaia EDR3 \citep{GaiaCollaboration2021}. Similar to \cite{Gandolfi2018}, we use the Gaia $\mathrm{G_{RP}}$ bandpass as an approximation to the TESS bandpass. Gaia EDR3 reveals there are no bright stellar companions in each aperture having $\Delta~\mathrm{G_{RP}}<4$, where $\Delta~\mathrm{G_{RP}}$ is the difference between the $\mathrm{G_{RP}}$ magnitude of a star and the respective value for the TOI host star. 

\begin{figure*}[!ht]
\epsscale{1.15}
\plotone{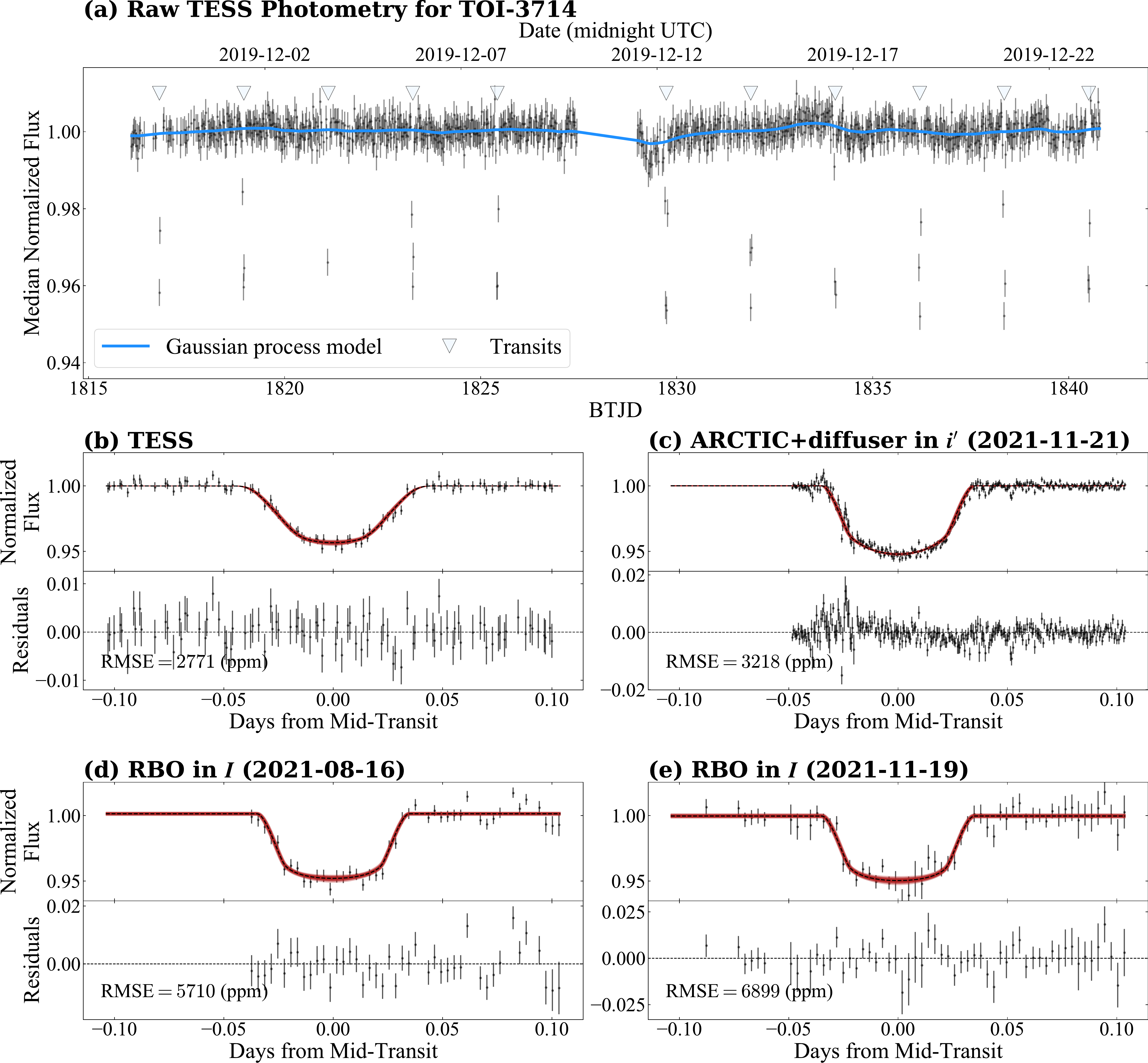}
\caption{\textbf{(a)} The median normalized TESS light curve for TOI-3714 derived with \texttt{eleanor}. The solid blue line is the best-fitting Gaussian process model used to detrend the light curve. The mid-transit times are indicated by the triangles. \textbf{(b)}$-$\textbf{(e)} are the light curves for TESS, ARCTIC, and RBO. In (b)-(e), the best-fitting model from the joint fit to the photometry and RVs is plotted as a dashed line while the shaded regions denote the \(1\sigma\) (darkest), \(2\sigma\), and \(3\sigma\) (lightest) extent of the model posteriors. The modeling of the photometry and RVs is described in detail in Section \ref{sec:modelfit}.}
\label{fig:3714phot}
\end{figure*}

\begin{figure*}[!ht]
\epsscale{1.15}
\plotone{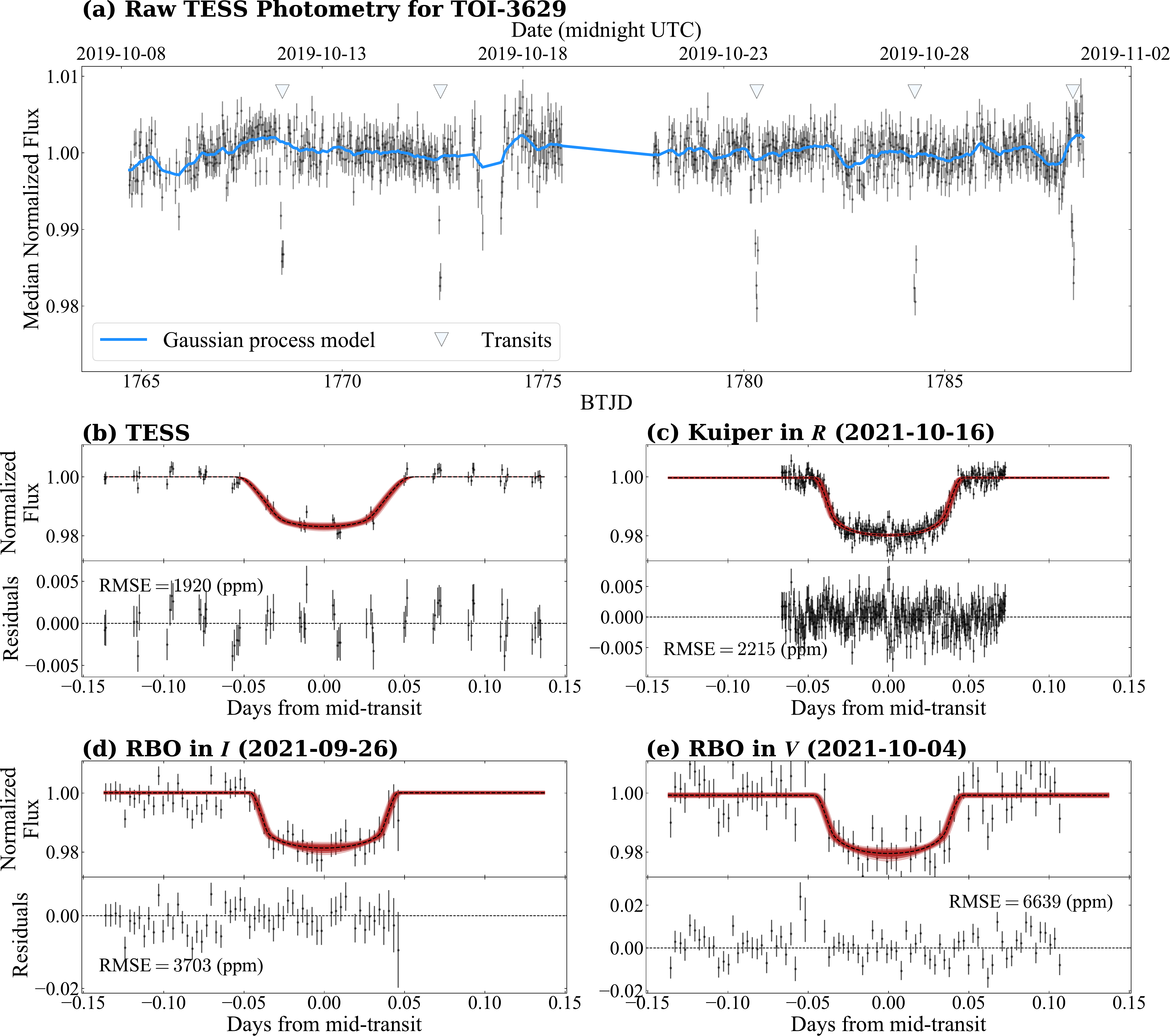}
\caption{Identical to Figure \ref{fig:3714phot}, but for TOI-3629. \textbf{(a)} The median normalized TESS light curve for TOI-3629 derived with \texttt{eleanor} along with the best-fitting Gaussian process model. The triangles indicate the mid-transit times. \textbf{(b)}$-$\textbf{(e)} are the light curves for the TESS, Kuiper, and RBO plotted with model posteriors (shaded regions) from the joint fit to the photometry and RVs.}
\label{fig:3629phot}
\end{figure*}

The TESS light curves used in this work are the \texttt{CORR\_FLUX} values calculated by \texttt{eleanor}. The corrected flux removes signals correlated with position (x and y pixel position), measured background, and time in the simple aperture flux. Observations where the background is larger than the stellar flux (\texttt{FLUX\_BKG} $>$ \texttt{CORR\_FLUX}) or with non-zero data quality flags \citep[Table 28 in][]{Tenenbaum2018} are excluded from analysis. Figures \ref{fig:3714phot} and \ref{fig:3629phot} present all photometry, including the TESS light curve, analyzed in this work.

\subsection{RBO 0.6m Telescope}
We used the 0.6m telescope at the Red Buttes Observatory (RBO) in Wyoming \citep{Kasper2016} to observe (i) TOI-3714 on the nights of 2021 August 16 and 2021 November 19 and (ii) TOI-3629 on the nights of 2021 September 26 and 2021 October 4. The 0.6m telescope is a f/8.43 Ritchey-Chr\'etien Cassegrain constructed by DFM Engineering, Inc. and equipped with an Apogee Alta F16M camera. The observations on 2021 October 4 of TOI-3629 were obtained in the Bessell V filter \citep{Bessell1990} while the other observations were obtained in the Bessell I filter. All observations used the $2 \times 2$ on-chip binning mode, which has a gain of 1.39 $\mathrm{e-/ADU}$, a plate scale of \(0.731 \arcsec/\mathrm{pixel}\), and a readout time of $\sim2.4$s. Each target was defocused moderately and observed using an exposure time of 240s.

The RBO light curves were derived using \texttt{AstroImageJ} \citep{Collins2017}. Following the methodology in \cite{Stefansson2017}, the estimated scintillation noise was included in the flux uncertainty. The final reductions used a photometric aperture radius of 10 pixels ($7.3\arcsec$), an inner sky radius of 20 pixels ($14.6\arcsec$) and an outer sky radius of 30 pixels ($21.9\arcsec$).

\subsection{APO 3.5m Telescope}
We used the 3.5m Astrophysical Research Consortium (ARC) Telescope Imaging Camera \citep[ARCTIC;][]{Huehnerhoff2016} on the ARC 3.5m Telescope at Apache Point Observatory (APO) to obtain a transit of TOI-3714 on the night of 2021 November 21. The observations were performed in the Sloan $i^\prime$ filter using an engineered diffuser \citep{Stefansson2017} with an exposure time of 45s. The average seeing for the night was $\sim1.0\arcsec$. ARCTIC was operated in the quad and fast readout modes using the $4 \times 4$ on-chip binning mode to achieve a gain of 2 $\mathrm{e-/ADU}$, a plate scale of $0.456 \mathrm{\arcsec/pixel}$, and a readout time of 2.7 s. Similar to the RBO data, we processed the photometry using \texttt{AstroImageJ} and include the scintillation noise estimate in the flux uncertainty. The final reduction used a photometric aperture radius of 10 pixels ($4.6\arcsec$), an inner sky radius of 20 pixels ($9.1\arcsec$) and outer sky radius of 30 pixels ($13.7\arcsec$).

\subsection{Kuiper 61'' Telescope}
We used the 61'' (1.55m) Kuiper Telescope located on Mt. Bigelow, Arizona to observe TOI-3629 on the night of 2021 October 16. The Kuiper Telescope\footnote{\url{http://james.as.arizona.edu/~psmith/61inch/CCD/basicinfo.html}} is equipped with the Mont4k imager, which uses a $4096\times4097$ Fairchild CCD486 detector to provide a field of view of $9.7\arcmin\times9.7\arcmin$. TOI-3629 was observed in the Harris R-band using a 30 s exposure time with an average seeing of $\sim1.7\arcsec$. The pixels were binned in $3\times3$ mode to shorten the readout time. This achieves a plate scale of 0.42\arcsec/pixel. Similar to the RBO data, we processed the photometry using \texttt{AstroImageJ} and include the scintillation noise estimate in the flux uncertainty. The final reduction used a photometric aperture radius of 8 pixels ($3.6\arcsec$), an inner sky radius of 14 pixels ($6.3\arcsec$) and outer sky radius of 22 pixels ($9.9\arcsec$).

\subsection{ZTF photometry}
ZTF data for TOI-3714 and TOI-3629 are publicly available under DR11\footnote{\url{https://www.ztf.caltech.edu/ztf-public-releases.html}}. Both objects were observed through a public program designed to observe TESS northern sectors by ZTF \citep{vanRoestel2019}. ZTF has a plate scale of $1.012\arcsec{}\mathrm{~pixel^{-1}}$ \citep{Yao2019} and the exposures for all observations are 30 s long. We follow the advice of the ZTF Science Data System Explanatory Supplement\footnote{\url{https://web.ipac.caltech.edu/staff/fmasci/ztf/ztf_pipelines_deliverables.pdf}} (ZDS) and reject bad quality data with (i) non-zero \texttt{catflag} values (see \textsection 13.6 in ZDS), (ii) values of $\chi\ge4$, where $\chi$ is the rms of the residuals to the PSF fit on the source performed by the ZTF pipeline, and (iii) values of $|\mathtt{sharp}|\ge0.5$, where \texttt{sharp} is the difference of the observed and model squared PSF FWHM. TOI-3714 has (i) 512 observations spanning 2018 April 08 through 2022 March 2 with a median cadence of 1 day and a median precision of $\sim1.3\%$ in the $zr$ filter and (ii) 355 observations spanning 2018 March 29 through 2022 March 2 with a median cadence of 2 days and a median precision of $\sim1.7\%$ in the $zg$ filter. TOI-3629 has (i) 695 observations spanning 2018 May 18 through 2022 February 18 with a median cadence of 1 day and a median precision of $\sim1.0\%$ in the $zr$ filter and (ii) 574 observations spanning 2018 May 25 through 2022 February 18 with a median cadence of 1 day and a median precision of $\sim1.1\%$ in the $zg$ filter. 

\subsection{High-contrast imaging}
TOI-3629 and TOI-3714 were observed on 2021 October 25 and 2021 December 21 respectively, using the speckle imaging instrument NESSI \citep{Scott2018} on the WIYN 3.5m Telescope at Kitt Peak National Observatory (KPNO). Due to the faintness of these targets ($r^\prime>14$), the images were acquired in Sloan \(r^\prime\) (TOI-3629 only) and \(z^\prime\) instead of the narrower filters that NESSI traditionally uses. TOI-3714 was observed only in the Sloan \(z^\prime\) filter because hardware issues during the observing run allowed for operations only with the redder filter. The images in each filter were reconstructed following the procedures outlined in \cite{Howell2011}.

\begin{figure*}[!ht]
\epsscale{1.15}
\plotone{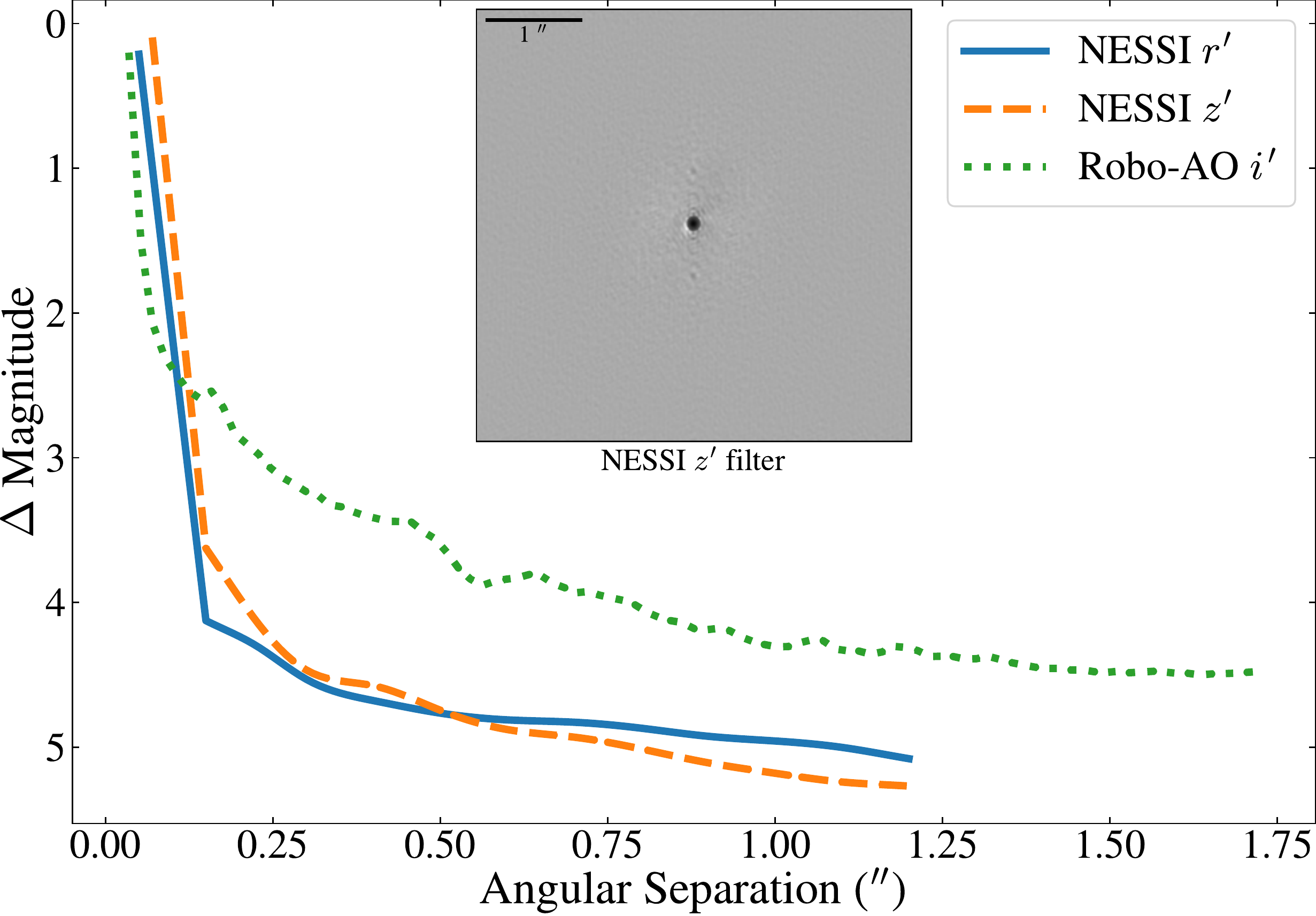}
\caption{The \(5\sigma\) contrast curves for TOI-3629 obtained from AO imaging using Robo-AO in the Sloan \(i^\prime\) filter and speckle imaging with NESSI in the Sloan \(r^\prime\) and \(z^\prime\) filters. The data reveal no bright companions at separations of $0.2\arcsec{}-1.75\arcsec{}$ from TOI-3629. The inset is the $4.7\arcsec{}\times4.7\arcsec{}$ NESSI speckle image centered on TOI-3629 in the Sloan \(z^\prime\) filter.}
\label{fig:3629imaging}
\end{figure*}

\begin{figure*}[!ht]
\epsscale{1.15}
\plotone{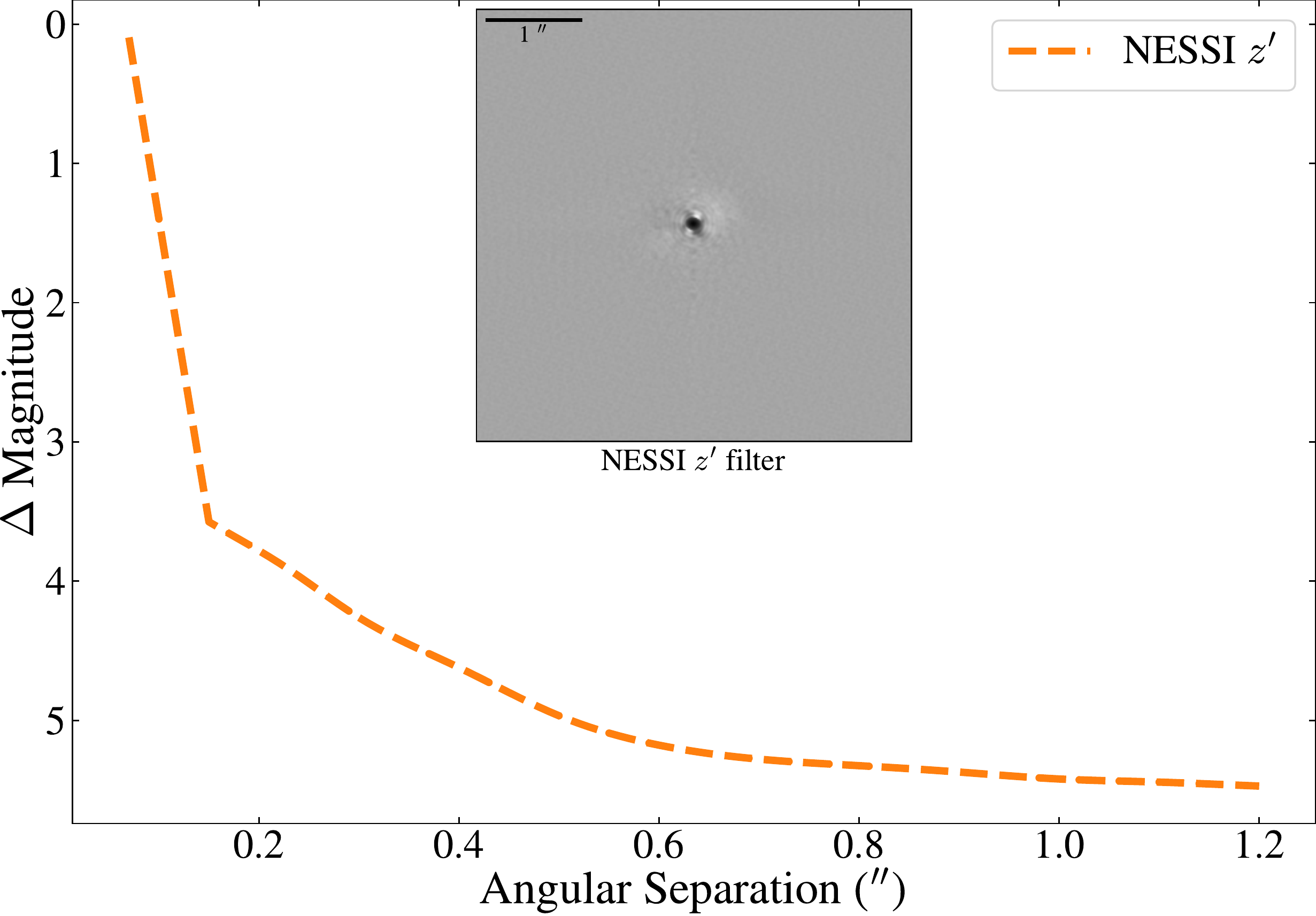}
\caption{Similar to Figure \ref{fig:3629imaging} but for TOI-3714.  The \(5\sigma\) contrast curve for TOI-3714 obtained from speckle imaging using NESSI in the Sloan \(z^\prime\) filter. The data reveal no bright companions at separations of $0.2\arcsec{}-1.75\arcsec{}$ from TOI-3714. The inset is the $4.7\arcsec{}\times4.7\arcsec{}$ NESSI speckle image centered on TOI-3714 in the Sloan \(z^\prime\) filter.}
\label{fig:3714imaging}
\end{figure*}

TOI-3629 was also observed as part of the Robo-AO Kepler M dwarf multiplicity survey \citep{Lamman2020} on 2016 October 19. The observations were performed using the Robo-AO laser adaptive optics system \citep{Baranec2013,Baranec2014} on the 2.1m telescope at KPNO \citep{Jensen-Clem2018} using a 1.85m circular aperture mask on the primary mirror. These observations were taken in the Sloan \(i^\prime\) filter. \cite{Lamman2020} have made the Robo-AO contrast curve for TOI-3629 publicly available on ExoFOP-TESS\footnote{\url{https://exofop.ipac.caltech.edu/tess/view_tag.php?tag=13940}}. The Robo-AO observation reveals no bright ($\Delta \mathrm{mag} < 4$) stellar companions at separations of $0.75-1.75\arcsec$ from TOI-3629.

Figures \ref{fig:3629imaging} and \ref{fig:3714imaging} show the $5\sigma$ contrast curves for TOI-3629 and TOI-3714, respectively, along with an inset of the NESSI speckle image in the $z^\prime$ filter. Together, the NESSI and Robo-AO data show there are no bright ($\Delta \mathrm{mag} < 4$) companions and no significant source of dilution at separations of $0.2-1.2\arcsec$ from either target. 

\subsection{HPF spectrograph}
HPF is a high-resolution ($R\sim55,000$), fiber-fed \citep{Kanodia2018a}, temperature controlled \citep{Stefansson2016},  near-infrared (\(\lambda\sim8080-12780\)\ \AA) spectrograph located on the 10m HET at McDonald Observatory in Texas \citep{Mahadevan2012,Mahadevan2014}. Observations are executed in a queue by the HET resident astronomers \citep{Shetrone2007}. Between 2021 January 18 and 2022 January 14, we obtained 12 visits of TOI-3714 and 23 visits of TOI-3629. The median signal-to-noise ratios (S/N) per 1D extracted pixel at 1070nm are 44 and 54, respectively, for these targets. 

The \texttt{HxRGproc} tool\footnote{\url{https://github.com/indiajoe/HxRGproc}} \citep{Ninan2018} was used to process the raw HPF data and perform bias noise removal, nonlinearity correction, cosmic-ray correction, and slope/flux and variance image calculation. The one-dimensional spectra were extracted following the procedures in \cite{Ninan2018}, \cite{Kaplan2019}, and \cite{Metcalf2019}. The wavelength solution and drift correction were extrapolated using laser frequency comb (LFC) frames obtained from routine calibrations. This extrapolation enables wavelength calibration on the order of $<30~\mathrm{cm~s^{-1}}$ \citep[see Appendix A in][]{Stefansson2020}, a value which is much smaller than the RV uncertainty for our targets (\(>10~\mathrm{m~s^{-1}}\)). 

The RVs were calculated using a modified version of the \texttt{SpEctrum Radial Velocity AnaLyser} code \citep[SERVAL;][]{Zechmeister2018} optimized for HPF RV extractions (see \cite{Metcalf2019} and \cite{Stefansson2020} for details). \texttt{SERVAL} employs the template-matching technique to derive RVs \citep[e.g.,][]{Anglada-Escude2012} and creates a master template from the observations to determine the Doppler shift by minimizing the \(\chi^2\) statistic. The master template is generated from all observed spectra after masking sky-emission lines and telluric regions identified using a synthetic telluric-line mask generated from \texttt{telfit} \citep{Gullikson2014}. The barycentric correction is calculated using \texttt{barycorrpy}, a Pythonic implementation \citep{Kanodia2018} of the algorithms from \cite{Wright2014}. 

\begin{deluxetable}{lrcccc}
{\tabletypesize{\scriptsize }
\tablecaption{RVs of TOI-3714 and TOI-3629. \label{tab:rvs}}
\tablehead{
\colhead{$\mathrm{BJD_{TDB}}$}  &
\colhead{RV} &
\colhead{$\sigma$} & 
\colhead{S/N$^a$} &
\colhead{Exp. Time} &
\colhead{Instrument}\\
& 
\colhead{$(\mathrm{m~s^{-1}})$} & 
\colhead{$(\mathrm{m~s^{-1}})$} & 
&
\colhead{(s)} & 
}
\startdata
\multicolumn{4}{l}{\hspace{-0.2cm} TOI-3714:}  \\
~~2459450.941268 &  $-196$ &  23 & 44 &  1890 & HPF \\
~~2459451.948741 &   163 &  25 & 42 &  1890 & HPF \\
~~2459452.941751 &  $-183$ &  22 & 47 &  1890 & HPF \\
~~2459458.924675 &    15 &  25 & 41 &  1890 & HPF \\
~~2459511.784103 &    60 &  23 & 44 &  1890 & HPF \\
~~2459512.783671 &    29 &  23 & 46 &  1890 & HPF \\
~~2459516.779359 &   209 &  26 & 40 &  1890 & HPF \\
~~2459516.995256 &    64 &  20 & 50 &  1890 & HPF \\
~~2459518.748189 &   119 &  30 & 35 &  1890 & HPF \\
~~2459518.992056 &   135 &  23 & 44 &  1890 & HPF \\
~~2459519.985625 &  $-189$ &  22 & 46 &  1890 & HPF \\
~~2459571.844384 &  $-103$ &  29 & 34 &  1890 & HPF \\
~~2459479.884140 &    71 &  12 & 15 &  1800  & NEID \\
~~2459503.998300 &    27 &  15 & 11 &  1200  & NEID \\
~~2459520.927772 &    77 &  11 & 15 &  1800  & NEID \\
~~2459531.844720 &    58 &  10 & 16 &  1800  & NEID \\
~~2459533.801149 &    89 &   9 & 18 &  1800  & NEID \\
~~2459560.766674 &  $-244$ &  14 & 12 &  1800  & NEID \\
~~2459586.625543 &  $-239$ &  11 & 15 &  1800  & NEID \\
~~2459587.851477 &    91 &  20 & 9 &  1800  & NEID \\
\hline
\multicolumn{4}{l}{\hspace{-0.2cm} TOI-3629:}  \\
~~2459232.579925 &   26 &  16 & 52 &  1890  & HPF \\
~~2459233.576944 &  $-24$ &  18 & 48 &  1890  & HPF \\
~~2459448.764453 &   28 &  13 & 66 &  1890  & HPF \\
~~2459451.979492 &    1 &  18 & 49 &  1890  & HPF \\
~~2459452.761162 &    5 &  15 & 59 &  1890  & HPF \\
~~2459453.979751 &  $-72$ &  16 & 58 &  1890  & HPF \\
~~2459455.739966 &    5 &  14 & 62 &  1890  & HPF \\
~~2459457.974338 &  $-42$ &  13 & 64 &  1890  & HPF \\
~~2459460.962548 &  $-16$ &  15 & 57 &  1890  & HPF \\
~~2459461.955253 &  $-91$ &  18 & 48 &  1890  & HPF \\
~~2459470.709176 &  $-47$ &  17 & 51 &  1890  & HPF \\
~~2459471.707553 &    9 &  14 & 60 &  1890  & HPF \\
~~2459475.919121 &   23 &  20 & 47 &  1890  & HPF \\
~~2459477.918199 &  $-72$ &  18 & 50 &  1890  & HPF \\
~~2459480.910306 &  $-10$ &  24 & 51 &   945  & HPF \\
~~2459485.896427 &  $-57$ &  22 & 41 &  1890  & HPF \\
~~2459499.627624 &   64 &  17 & 54 &  1890  & HPF \\
~~2459507.814595 &   24 &  23 & 40 &  1890  & HPF \\
~~2459516.581430 &  $-20$ &  15 & 59 &  1890  & HPF \\
~~2459543.736198 &   11 &  15 & 60 &  1890  & HPF \\
~~2459588.597139 &  $-57$ &  15 & 56 &  1890  & HPF \\
~~2459592.597117 &  $-48$ &  15 & 60 &  1890  & HPF \\
~~2459593.588844 &   44 &  16 & 54 &  1890  & HPF \\
~~2459478.965087 &  $-20$ &   6 & 22 &  1800  & NEID \\
~~2459479.794197 &   28 &   6 & 22 &  1800  & NEID \\
~~2459528.888224 &  $-34$ &  14 & 11 &  1800  & NEID \\
~~2459532.843589 &  $-38$ &   8 & 19 &  1800  & NEID \\
~~2459546.840787 &   62 &  15 & 11 &  1800  & NEID \\
\enddata
\tablenotetext{a}{The HPF and NEID S/N are the median values per 1D extracted pixel at 1070nm and 850nm, respectively.}
}
\end{deluxetable}

\subsection{NEID spectrograph}
NEID is an environmentally stabilized \citep{Stefansson2016,Robertson2019}, high-resolution ($R\sim110,000$) spectrograph installed on the WIYN 3.5m telescope at KPNO in Arizona \citep[][]{Schwab2016}. NEID features extended red wavelength coverage (\(\lambda\sim3800-9300\)\ \AA) and a fiber-feed system similar to HPF \citep{Kanodia2018a}. Between 2021 September 21 and 2022 January 8, we obtained 8 visits of TOI-3714 and 5 visits of TOI-3629. Observations were obtained in queue mode and NEID operated in high-resolution mode. The median S/N per 1D extracted pixel was 15 and 19, respectively, at 850nm. 

\begin{figure*}[!ht]
\epsscale{1.15}
\plotone{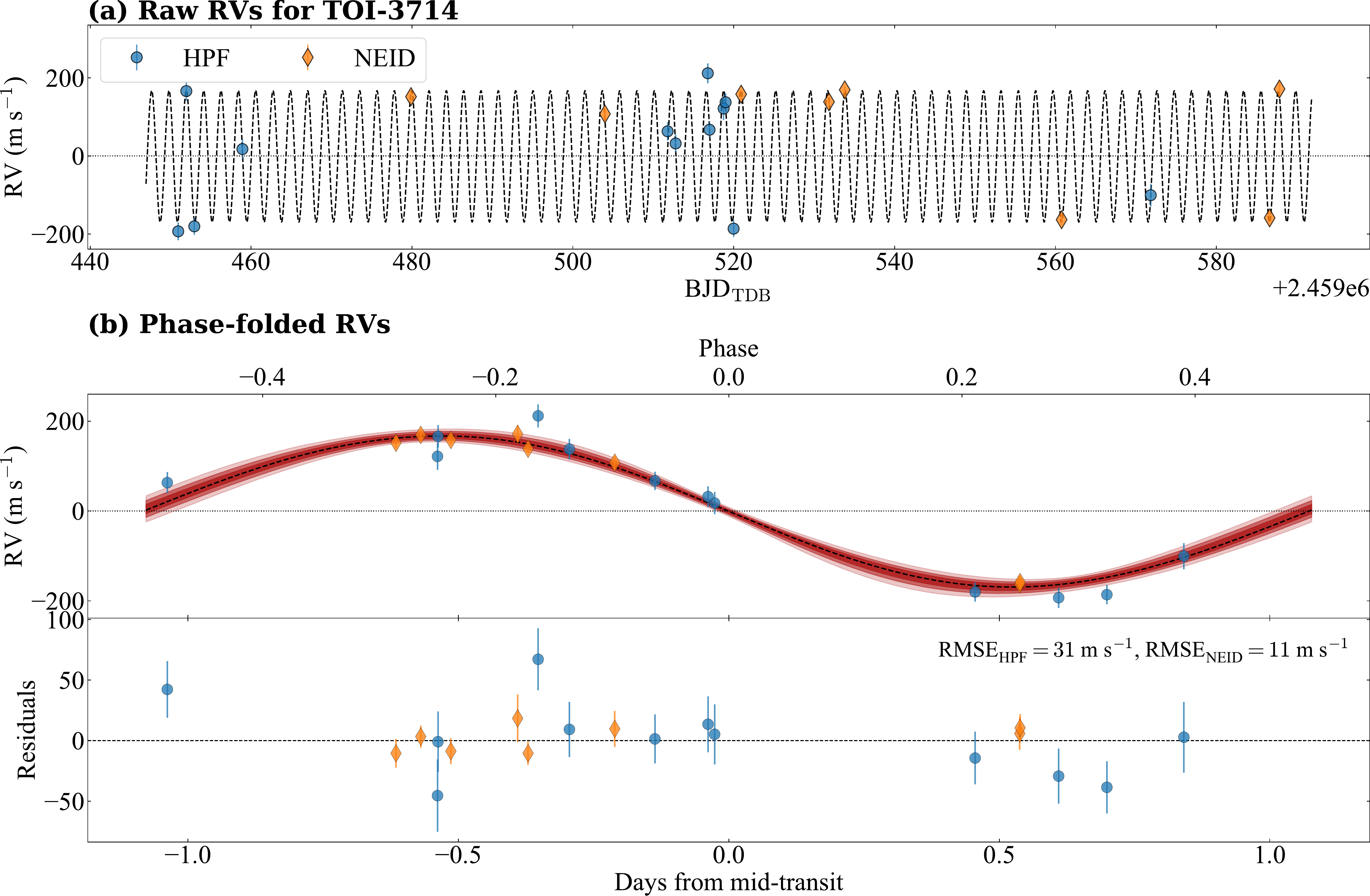}
\caption{\textbf{(a)} shows the RVs for TOI-3714, after subtracting the instrumental offsets, derived with modified versions of \texttt{SERVAL}. \textbf{(b)} displays the phase-folded RVs plotted with model posteriors. For each panel, the dashed line is the best-fitting Keplerian model. The shaded regions denote the \(1\sigma\) (darkest), \(2\sigma\), and \(3\sigma\) (lightest) extent of the model posteriors. The modeling is described in Section \ref{sec:modelfit}.}
\label{fig:3714rv}
\end{figure*}

\begin{figure*}[!ht]
\epsscale{1.15}
\plotone{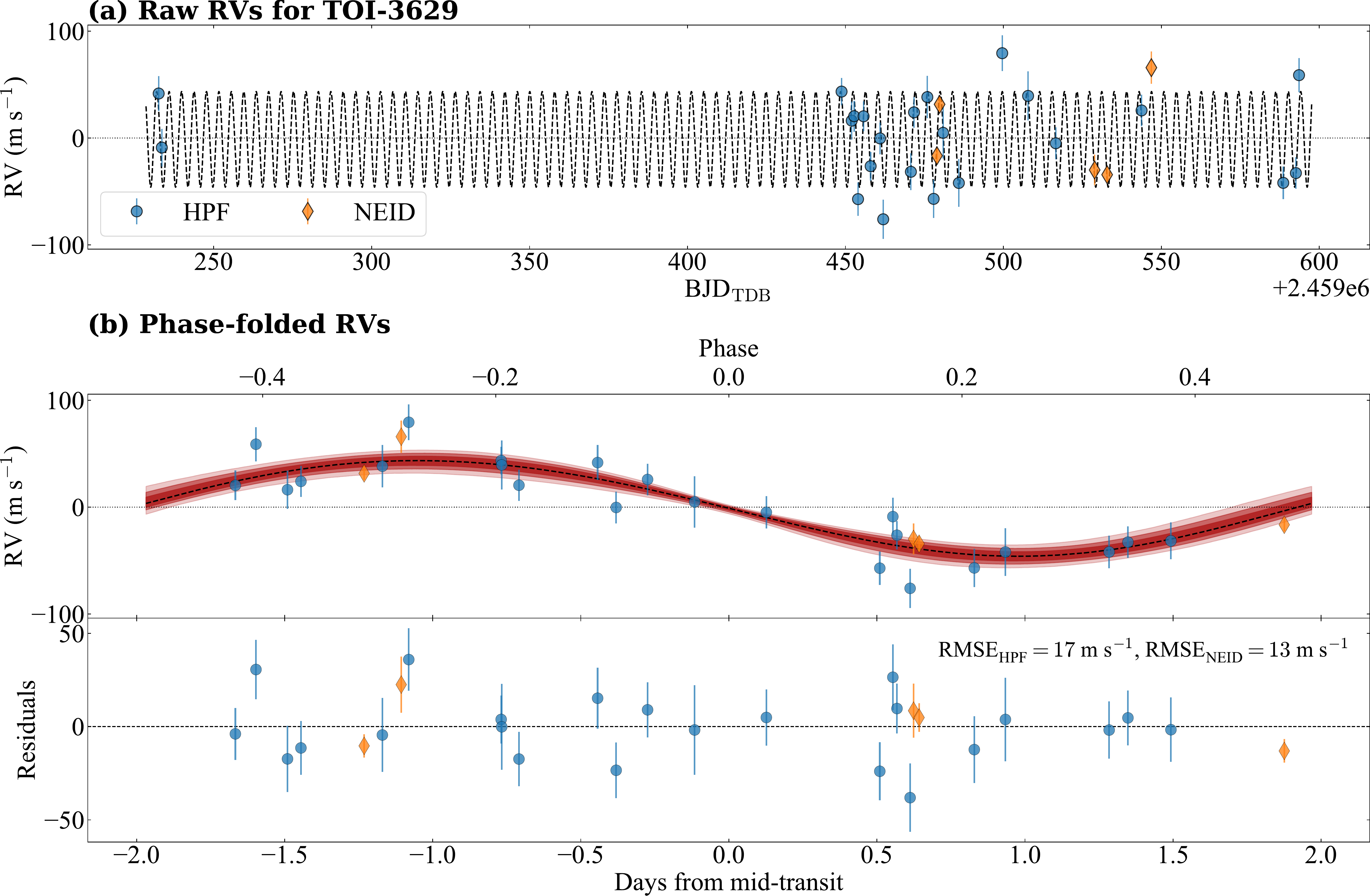}
\caption{Identical to Figure \ref{fig:3714rv}, but for TOI-3629. \textbf{(a)} shows RVs for TOI-3629, after subtracting the instrumental offsets. \textbf{(b)} displays the phase-folded RVs plotted with model posteriors.}
\label{fig:3629rv}
\end{figure*}

The NEID data were reduced using the NEID Data Reduction Pipeline\footnote{\url{https://neid.ipac.caltech.edu/docs/NEID-DRP/}} (DRP), and the Level-2 1D extracted spectra were retrieved from the NEID Archive\footnote{\url{https://neid.ipac.caltech.edu/}}. Similar to HPF, to maximize the RV precision from the M dwarf spectra, we measured the RVs using a modified version of the \texttt{SERVAL} code \cite[see][]{Stefansson2021}. We extracted RVs with \texttt{SERVAL} using different segments of an order (inner 3000, 5000, or 7000 pixels) and different wavelength ranges ($4950-8960$ \AA{} or $5440-8960$ \AA{}). The various combinations of pixel and wavelength ranges produced RVs that, when jointly modeled with photometry and HPF RVs, resulted in identical system parameters (within their $1\sigma$ uncertainty). The NEID RVs presented in this work were calculated using the wavelength range from $5440-8920$ \AA{} (order indices $61-104$) and the inner most 3000 pixels of each order. This effectively uses the central blaze region of each order and limits the use of the lower S/N regions near the edge of each order. Table \ref{tab:rvs} reports the HPF and NEID RVs, the \(1\sigma\) uncertainties, the S/N per pixel, and exposure times for TOI-3714 and TOI-3629. Figures \ref{fig:3714rv} and \ref{fig:3629rv} display the RVs for TOI-3714 and TOI-3629, respectively.

\section{Stellar Parameters}\label{sec:stellarpar}
\subsection{Spectroscopic parameters}\label{sec:specmatch}
The stellar effective temperature ($T_e$), surface gravity ($\log g_\star$), and metallicity ([Fe/H]) were calculated using the \texttt{HPF-SpecMatch}\footnote{\url{https://gummiks.github.io/hpfspecmatch/}} package \citep[][]{Stefansson2020}, which derives stellar parameters using the empirical template matching methodology discussed in \cite{Yee2017}. It identifies the best-matching spectra from a library of well-characterized stars using \(\chi^{2}\) minimization, creates a composite spectrum using a weighted, linear combination of the five best-matching library spectra, and derives the stellar properties using these weights. When searching for the best-matching library spectra, \texttt{HPF-SpecMatch} broadens the stellar templates using a linear limb darkening law. The reported uncertainty is the standard deviation of the residuals from a leave-one-out cross-validation procedure applied to the entire spectral library in the chosen spectral order.

The HPF spectral library contains 166 stars and spans the following parameter space: $2700~\mathrm{K} < T_{e} < 6000~\mathrm{K}$, $4.3<\log g_\star < 5.3$, and $-0.5 < \mathrm{[Fe/H]} < 0.5$. The library includes 87 M dwarfs ($T_{e}\le4000$ K) of which 37 are early M dwarfs spanning $3500\mathrm{K} \le T_{e} \le 4000~\mathrm{K}$, $4.6<\log g_\star < 4.9$, and $-0.5 < \mathrm{[Fe/H]} < 0.5$. The spectral matching was performed on HPF order index 5 ($8534-8645$ \AA) for both targets because this order has little to no telluric contamination. The resolution limit of HPF places a constraint of $v \sin i < 2 \mathrm{~km~s^{-1}}$ for both TOI-3714 and TOI-3629. TOI-3714 is determined to have $T_{e}=3660\pm90$ K, $\log g_\star=4.75\pm0.05$, and $\mathrm{[Fe/H]=0.1\pm0.1}$. TOI-3629 is determined to have $T_{e}=3870\pm90$ K, $\log g_\star=4.67\pm0.05$, and $\mathrm{[Fe/H]=0.4\pm0.1}$. Table \ref{tab:stellarparam} presents the derived spectroscopic parameters with their uncertainties. 

\begin{deluxetable*}{lcccc}
{\tabletypesize{\tiny}
\rotate
\tablecaption{Summary of Stellar Parameters. \label{tab:stellarparam}}
\tablehead{\colhead{~~~Parameter}&  \colhead{Description}&
\colhead{TOI-3714}&
\colhead{TOI-3629}&
\colhead{Reference}}
\startdata
\multicolumn{4}{l}{\hspace{-0.2cm} Main Identifiers:}  \\
~~~TIC &  \(\cdots\)  & 155867025 & 455784423 & TIC \\
~~~Gaia EDR3 & \(\cdots\) & 178924390478792320 & 2881820324294985856 & Gaia EDR3 \\
\hline
\multicolumn{4}{l}{\hspace{-0.2cm} Coordinates, Proper Motion, Distance, Maximum Extinction, and Spectral Type:} \\
~~~$\alpha_{\mathrm{J2016}}$ &  Right Ascension (RA) & 04:38:12.56 & 23:59:10.42 & Gaia EDR3 \\
~~~$\delta_{\mathrm{J2016}}$ &  Declination (Dec) & 39:27:28.77 & 39:18:51.32 & Gaia EDR3 \\
~~~$\mu_{\alpha}$ &  Proper motion (RA, mas yr$^{-1}$) & $19.83 \pm 0.03$ & $185.71 \pm 0.01$ & Gaia EDR3 \\
~~~$\mu_{\delta}$ &  Proper motion (Dec, mas yr$^{-1}$) & $-70.74 \pm 0.02$ & $1.01 \pm 0.01$ & Gaia EDR3  \\
~~~$l$ &  Galactic longitude & 163.30437 & $112.02292$ & Gaia EDR3 \\
~~~$b$ &  Galactic latitude & $-5.02268$ & $-22.44783$ & Gaia EDR3  \\
~~~$d$ &  Geometric distance (pc)  & $112.5_{-0.4}^{+0.2}$ & $129.7 \pm 0.3$ & Bailer-Jones\\
~~~\(A_{V,max}\) & Maximum visual extinction & $0.02$ & $0.01$ & Green\\
~~~Spectral Type & \(\cdots\) & M$2\pm0.5$ & M$1\pm0.5$ & LAMOST \\
\hline
\multicolumn{4}{l}{\hspace{-0.2cm} Broadband Photometric Magnitudes:}  \\
~~~$B$ & Johnson $B$ mag & $16.8 \pm 0.2$ & $16.1 \pm 0.1$ & APASS\\
~~~$V$ & Johnson $V$ mag & $15.24 \pm 0.09$ & $14.63 \pm 0.05$ & APASS\\
~~~$g'$ & Sloan $g'$ mag & $15.9 \pm 0.1$ & $15.33 \pm 0.04$ & APASS\\
~~~$r'$ & Sloan $r'$ mag & $14.73 \pm 0.09$ & $14.06 \pm 0.05$ & APASS\\
~~~$i'$ & Sloan $i'$ mag & $13.66 \pm 0.09$ & $13.12 \pm 0.07$ & APASS\\
~~~$J$ & $J$ mag & $11.74 \pm 0.02$ & $11.42 \pm 0.03$ & 2MASS\\
~~~$H$ & $H$ mag & $11.06 \pm 0.02$ & $10.73 \pm 0.03$ & 2MASS\\
~~~$K_s$ & $K_s$ mag & $10.85 \pm 0.02$ & $10.55 \pm 0.02$ & 2MASS\\
~~~W1 & WISE1 mag & $10.72 \pm 0.02$ & $10.48 \pm 0.02$ & WISE\\
~~~W2 &  WISE2 mag & $10.68 \pm 0.02$ & $10.52 \pm 0.02$ & WISE\\
~~~W3 &  WISE3 mag & $10.5 \pm 0.1$ & $10.38 \pm 0.07$ & WISE\\
\hline
\multicolumn{4}{l}{\hspace{-0.2cm} Spectroscopic Parameters$^a$:}\\
~~~$T_{e}$ &  Effective temperature (K) & $3660 \pm 90$ & $3870 \pm 90$& This work\\
~~~$\log g_\star$ & Surface gravity (cgs) & $4.75\pm0.05$ & $4.67 \pm 0.05$ & This work\\
~~~$\mathrm{[Fe/H]}$ &  Metallicity (dex) & $0.1\pm0.1$ & $0.4\pm0.1$ & This work\\
~~~$v\sin i_{\star}$ & Rotational broadening (km s$^{-1}$) & $<2$ & $<2$ & This work\\
\hline
\multicolumn{4}{l}{\hspace{-0.2cm} Model-dependent Stellar SED and Isochrone Fit Parameters$^b$:}\\
~~~$M_\star$ &  Mass ($M_{\odot}$) & $0.53 \pm 0.02$ & $0.63 \pm 0.02$ & This work\\
~~~$R_\star$ &  Radius ($R_{\odot}$) & $0.51 \pm 0.01$ & $0.60_{-0.01}^{+0.02}$ & This work\\
~~~$\rho_\star$ &  Density ($\mathrm{g~cm^{-3}}$) & $5.8_{-0.3}^{+0.4}$ & $4.0 \pm 0.2$ & This work\\
~~~$A_v$ & Visual extinction (mag) & $0.011 \pm 0.007$ & $0.005 \pm 0.003$ & This work\\
~~~Age$^c$ & Age (Gyrs) & $0.7-5.1$ & $7\pm2$ & This work\\
\hline
\multicolumn{4}{l}{\hspace{-0.2cm} Other Stellar Parameters:}\\
~~~$RV$ & Systemic RV (km s$^{-1}$) & $36.4 \pm 0.2$ & $-24.83 \pm 0.06$ & This work\\
~~~$P_{rot}$ & Stellar rotation period (days) & $23.3 \pm 0.3$ & $\cdots$ & This work\\
~~~$U, V, W$ &  Barycentric Galactic velocities (km s$^{-1}$) &  $-43.5 \pm 0.2, -23.9 \pm 0.1, -20.43 \pm 0.05$ & $-91.9 \pm 0.2, -72.0 \pm 0.1, -13.06 \pm 0.06$ & This work\\
~~~$(U, V, W)_{\mathrm{LSR}}$ &  Galactic velocities w.r.t. LSR$^d$ (km s$^{-1}$) &  $-32.4 \pm 0.8, -11.7 \pm 0.5, -13.2 \pm 0.4$ & $-80.8 \pm 0.8, -59.7 \pm 0.5, -5.8 \pm 0.4$ & This work\\
\enddata
\tablerefs{TIC \citep{Stassun2019}, Gaia EDR3 \citep{GaiaCollaboration2021}, Bailer-Jones \citep{Bailer-Jones2021}, Green \citep{Green2019}, LAMOST \citep{Zhong2019}, APASS \citep{Henden2018}, 2MASS \citep{Cutri2003}, WISE \citep{Wright2010}}
\tablenotetext{a}{Derived with the \texttt{HPF-SpecMatch} package.}
\tablenotetext{b}{Derived with the \texttt{EXOFASTv2} package using MIST isochrones.}
\tablenotetext{c}{We report the age estimate of TOI-3714 using the rotation period and classification from \cite{Newton2016}. The age for TOI-3629 is from Galactic models based on its tangential velocity.}
\tablenotetext{d}{Calculated using the solar velocities from \cite{Schoenrich2010}.}
}
\end{deluxetable*}

\subsection{Spectral classification}
The Large Sky Area Multi-Object Fibre Spectroscopic Telescope (LAMOST) collaboration observed TOI-3714 on 2012 January 11 and TOI-3629 on 2018 November 3 as part of a survey of the Galactic anti-center \citep{Yuan2015,Xiang2017}. LAMOST is a 4m telescope equipped with 4000 fibers distributed over a 5\degr\ FOV that is capable of acquiring spectra in the optical band (3700-9000\AA) at a resolution \(R\approx1800\) with a limiting magnitude of SDSS \(r^\prime=19\) mag \citep{Cui2012}. The data used in this work are from the public DR7v2.0\footnote{\url{http://dr7.lamost.org/}} release.  

The LAMOST stellar classification pipeline \citep[][]{Zhong2015} uses stellar templates to identify molecular absorption features (e.g., CaH, TiO) that are typical for M-type stars and reports the subclass of an M dwarf with an accuracy of $\pm0.5$ subtypes. To be classified as M dwarfs, targets must have (i) a mean S/N$>5$, (ii) a best-matching template that is an M type, and (iii) the spectral indices of the absorption features must be located in the M-type stellar regime identified in \cite{Zhong2019} ($0<\mathrm{TiO5}< 1.2$ and $0.6<\mathrm{CaH2+CaH3}< 2.4$). LAMOST classifies TOI-3714 as an M$2\pm0.5$ dwarf and TOI-3629 as an M$1\pm0.5$ dwarf, which agrees with the derived parameters from \texttt{HPF-SpecMatch} in Section \ref{sec:specmatch}.

\subsection{Spectral energy distribution fitting}
To derive model-dependent stellar parameters, we modeled the spectral energy distribution (SED) for each target using the {\tt EXOFASTv2} analysis package \citep{Eastman2019}. {\tt EXOFASTv2} calculates the bolometric corrections for the SED fit by linearly interpolating the precomputed bolometric corrections\footnote{\url{http://waps.cfa.harvard.edu/MIST/model_grids.html\#bolometric}} in \(\log g_\star\), \(\mathrm{T_{e}}\), [Fe/H], and \(A_V\) from the MIST model grids \citep{Dotter2016,Choi2016}. 

The SED fits use Gaussian priors on the (i) 2MASS \(J,~H,~K\) magnitudes, Sloan \(g^\prime,~r^\prime,~i^\prime\) magnitudes and Johnson \(B,~V\) magnitudes from \cite{Henden2018}, and Wide-field Infrared Survey Explorer magnitudes \citep{Wright2010}; (ii) $\log g_\star$, $T_{e}$, and [Fe/H] derived from \texttt{HPF-SpecMatch}, and (iii) the geometric distance calculated from \cite{Bailer-Jones2021} for each respective star. We apply an upper limit to the visual extinction based on estimates of Galactic dust \citep{Green2019} calculated at the distance determined by \cite{Bailer-Jones2021}. The \(R_{v}=3.1\) reddening law from \cite{Fitzpatrick1999} is used to convert the extinction from \cite{Green2019} to a visual magnitude extinction. Table \ref{tab:stellarparam} contains the stellar priors and derived stellar parameters with their uncertainties. The model-dependent mass and radius are (i) \(0.53\pm0.02~\mathrm{M_{\odot}}\) and \(0.51\pm0.01~\mathrm{R_{\odot}}\) for TOI-3714 and (ii) \(0.63\pm0.02~\mathrm{M_{\odot}}\) and \(0.60_{-0.01}^{+0.02}~\mathrm{R_{\odot}}\) for TOI-3629. The masses and radii are identical within their respective $1\sigma$ uncertainties to the parameters from the TIC catalog for TOI-3714 (\(0.51\pm0.02~\mathrm{M_{\odot}}\) and \(0.51\pm0.02~\mathrm{R_{\odot}}\)) and TOI-3629 (\(0.60\pm0.02~\mathrm{M_{\odot}}\) and \(0.61\pm0.02~\mathrm{R_{\odot}}\)).

\subsection{Stellar rotation period} \label{sec:prot}
If we assume TOI-3714 and TOI-3629 are well-aligned with the orbit of their transiting planets ($\sin i_\star\sim1$), the constraint of $v \sin i_\star<2\mathrm{~km~s^{-1}}$ from HPF spectra requires each star to have a rotation period of $>10$ days. We do not search for photometric modulation in the \texttt{corr\_flux} because long-period ($>10$ day) astrophysical signals, such as starspot-induced photometric variability, are attenuated when removing systematics with \texttt{eleanor} \citep{Feinstein2019}, similar to how long-period rotation signals are damped and distorted in Kepler PDCSAP light curves \cite{Gilliland2015,VanCleve2016}. We instead searched for photometric modulation in TESS data using the \texttt{TESS-SIP} package \citep{Hedges2020}, which is designed to simultaneously create a Lomb-Scargle periodogram and detrend systematics. \texttt{TESS-SIP} uses a linear model with two components: (i) regressors (by default 2 principal components and a mean offset) to remove instrument systematics and (ii) a sinusoidal component to fit a power spectrum. Only one sector of TESS data exists for each target and we limit this search to a rotation period range of $1-30$ days. For this search, the transits were excised using the duration and ephemeris from the QLP. \texttt{TESS-SIP} recovers no significant period for either TOI-3714 and TOI-3629 between $1-30$ days. 

We also used data from ZTF DR11 in the $zg$ and $zr$ filters to search for any long-period signals caused by activity-induced photometric modulations in the target stars. This search used the generalized Lomb-Scargle (GLS) periodogram \citep{Zechmeister2009} because it has been shown to successfully recover rotation periods in photometry \citep[see][]{VanderPlas2018,CantoMartins2020,Reinhold2020}. The GLS periodogram is based on Fourier decomposition and provides peaks in frequency space where the highest peak in the periodogram is associated with the period of the best fit sine wave that minimizes the $\chi^2$ statistic. We use the \texttt{GLS}\footnote{\url{https://github.com/mzechmeister/GLS}} package to perform this analysis and only consider significant peaks where the false alarm probability (FAP), as calculated following \cite{Zechmeister2009}, is below a threshold of $0.1\%$. Data within transits were excised using the duration and ephemeris from the QLP. A significant peak (FAP$<0.1\%$) of $\sim23.6$ days was found in both the $zr$ and $zg$ photometry of TOI-3714. A significant peak (FAP$<0.1\%$) of $\sim29.5$ days was found in the $zr$ photometry while no significant peaks were seen in the $zg$ photometry for TOI-3629.

To derive the rotation period and an estimate of its uncertainty, we modeled the ZTF photometry using the \texttt{juliet} analysis package \citep{Espinoza2019}, which performs the parameter estimation using the dynamic nested-sampling algorithm \texttt{dynesty} \citep{Speagle2020}. The photometric model is a Gaussian process noise model using the approximate quasi-periodic covariance function presented in \cite{Foreman-Mackey2017} of the form:
\begin{equation}
\resizebox{.88\hsize}{!}{$k(\tau) = \frac{B}{2 + C} e^{-\tau/L} \left[ \cos \left( \frac{2 \pi \tau}{P_\mathrm{GP}} \right) + (1 + C) \right],$}
 \label{eq:kernelperiodic}
\end{equation}
where $\tau$ is the time of observation while $B$, $C$, $L$, and $P_{\mathrm{GP}}$ are the hyperparameters of the covariance function. $B$ and $C$ represent the weight of the exponential term with a decay constant of $L$ (in days). $P_{\mathrm{GP}}$ determines the periodicity of the quasi-periodic oscillations, which is interpreted as the stellar rotation period. This kernel is able to reproduce the behavior of a more traditional quasi-periodic covariance function and has allowed for computationally efficient inference of stellar rotation periods even for large datasets that are not uniformly sampled \citep[e.g.,][]{Angus2018}. The photometric model includes a simple white-noise model $\sigma_{\mathrm{phot}}$ in the form of a jitter term that is added in quadrature to the error bars of each filter.

\begin{figure*}[!htb]
\epsscale{1.15}
\plotone{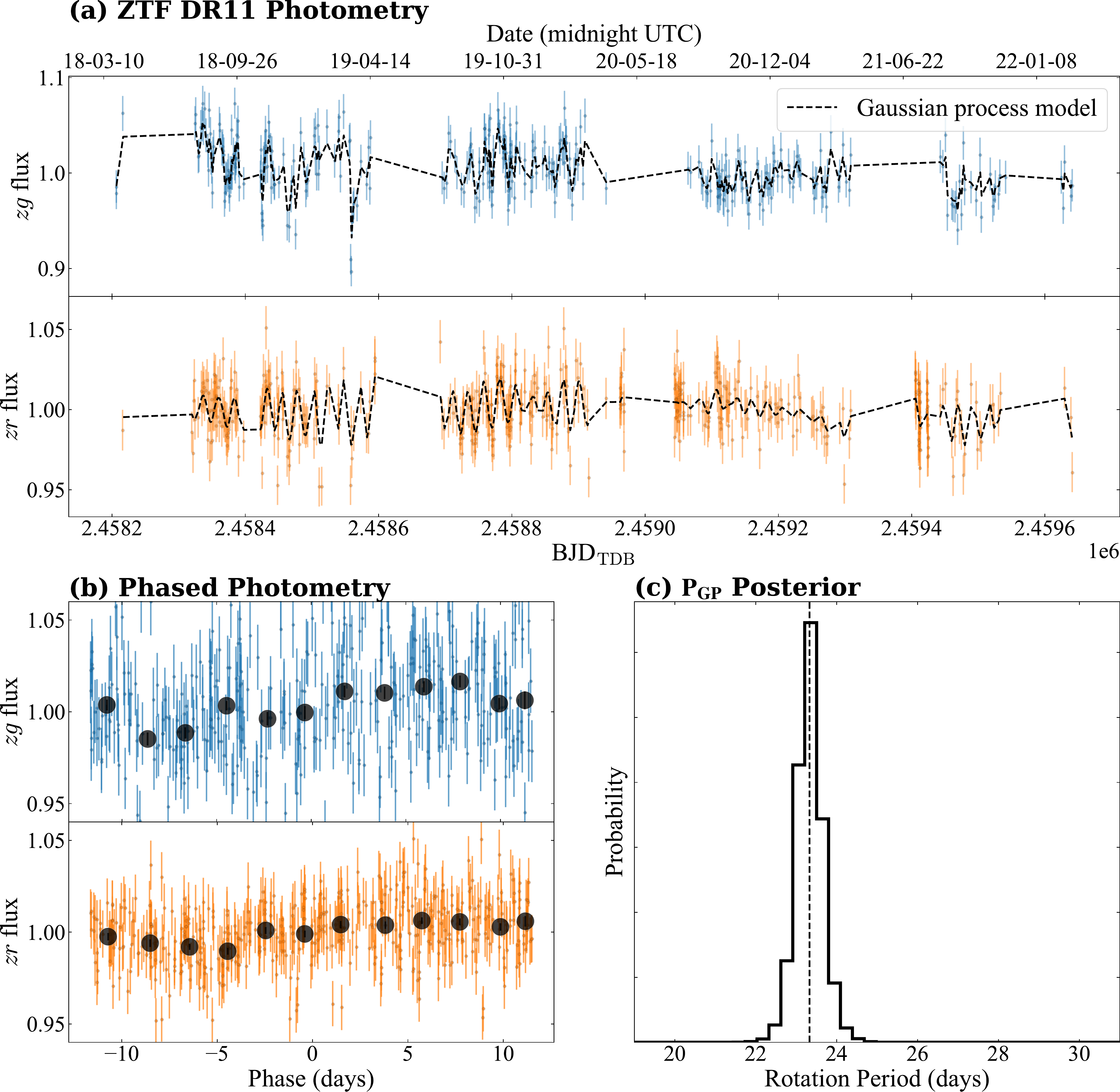}
\caption{\textbf{(a)} displays the ZTF photometry for TOI-3714 in each filter along with the best-fitting Gaussian process model for reference. \textbf{(b)} presents the ground-based ZTF photometry from panel (a) phased to the derived rotation period. The large black points represent 2-day bins of the phased photometry. \textbf{(c)} presents the posterior distribution of the rotation period from the Gaussian process model. The derived rotation period is \(23.3 \pm 0.3\) days.}
\label{fig:3714phasedrot}
\end{figure*}

The fit for each star uses a uniform prior on the Gaussian process period of $1-1500$ days where the upper limit coincides with the baseline of existing ZTF data.  For TOI-3714, the $zr$ and $zg$ data are jointly modeled and share the value of $P_{\mathrm{GP}}$, while the nuisance parameters ($B$, $C$, $L$, and $\sigma_\mathrm{phot}$) are different for each filter. For TOI-3629, we only model the $zr$ photometry. Figure \ref{fig:3714phasedrot} displays the ZTF photometry folded to the median value of $P_{\mathrm{GP}}$ and the posterior distribution for $P_{\mathrm{GP}}$, which we interpret as the rotation period. The measured rotation period is \(23.3 \pm 0.3\) days, which suggests that TOI-3714 most likely has an age between $0.7-5.1$ Gyr after adopting the classification scheme of \cite{Newton2016}. This age range is consistent with the rotation period and age relationship from \cite{Engle2018} and the values from our model-dependent SED fit. We are unable to place any additional constraint on the rotation period of TOI-3629 as the fit does not recover a significant period ($P_{\mathrm{GP}}=750_{-460}^{+450}$ days).

\subsection{Galactic kinematics}
The \textit{UVW} velocities in the barycentric frame were derived with \texttt{galpy} \citep{Bovy2015} using the Gaia EDR3 proper motions and the systemic velocity derived from HPF. The values in Table \ref{tab:stellarparam} are in a right-handed coordinate system \citep{Johnson1987} where \textit{UVW} are positive in the directions of the Galactic center, Galactic rotation, and the north Galactic pole, respectively. The \(UVW\) velocities in Table \ref{tab:stellarparam} are also provided with respect to the local standard of rest using the solar velocities and uncertainties from \cite{Schoenrich2010}. The BANYAN \(\Sigma\) algorithm \citep{Gagne2018}, which uses sky positions, proper motions, parallax, and RVs to constrain cluster membership probabilities, classifies both TOI-3714 and TOI-3629 as field stars showing no associations with known young clusters. 

Using kinematic selection criteria from \cite{Bensby2003}, TOI-3714 is classified as a member of the thin disk ($\mathrm{P_{Thick}}/\mathrm{P_{Thin}}=0.02$). The classification for TOI-3629 is ambiguous as the relative probability of thick disk to thin disk is $\mathrm{P_{Thick}}/\mathrm{P_{Thin}}=1.1$. We note that TOI-3629 has a large Galactic tangential velocity ($|V_T|=100.7 \pm 0.5\mathrm{~km~s^{-1}}$) with respect to the local standard of rest. \cite{Hwang2020a} used Galactic models to calculate the age distribution for different tangential velocity bins and determined a star with a tangential velocity in the range $100-120\mathrm{~km~s^{-1}}$ has a $\sim55\%$ chance of belonging to the thick disk and an estimated age of $7\pm2$ Gyr. While we cannot unambiguously classify TOI-3629 as a thick disk star, its high tangential velocity suggests that its age is most likely $>5$ Gyr.

\section{Photometric and RV Modeling} \label{sec:modelfit}
We use the \texttt{juliet} analysis package to jointly model the TESS photometry, ground-based photometry, and velocimetry and perform the parameter estimation using \texttt{dynesty}. \texttt{juliet} models the RVs with a standard Keplerian RV curve generated from the \texttt{radvel} \citep{Fulton2018} package and models the light curves with a transit model generated from the \texttt{batman} package \citep{Kreidberg2015}. The limb-darkening parameters are sampled from uniform priors following the parameterization presented in \cite{Kipping2013b}. For the long-cadence TESS photometry, the transit model utilizes the supersampling option in \texttt{batman} with exposure times of 30 minutes and a supersampling factor of 30. The photometric model also includes a dilution factor, \(D\), for TESS representing the ratio between the out-of-transit flux of the host star to that of all stars within the photometric aperture. We include this term for TESS data because \texttt{eleanor} does not correct for dilution from nearby stars despite the large apertures adopted (see Figures \ref{fig:3714apertures} and \ref{fig:3629apertures}). Both the photometric and RV models include a simple white-noise model parameterized as a jitter term that is added in quadrature to the error bars of each data set. To account for correlated noise in the TESS light curves, each fit includes a Gaussian process noise model of the same form described in Section \ref{sec:prot}. 

Tables \ref{tab:3714par} and \ref{tab:3629par} provide a summary of the priors used for the fit along with the inferred system parameters and the confidence intervals ($16\mathrm{th}-84\mathrm{th}$ percentile) for TOI-3714 and TOI-3629, respectively. Figures \ref{fig:3714phot}, \ref{fig:3629phot}, \ref{fig:3714rv} and \ref{fig:3629rv} display the model posteriors for each system. The modeling reveals that (i) TOI-3714 b is a hot Jupiter ($M_{2}=0.70 \pm 0.03~\mathrm{M_J}$ and $R_{2}=1.01 \pm 0.03~\mathrm{R_J}$) orbiting its host star on a nearly circular orbit with a period of \(2.154849 \pm 0.000001\) days and (ii) TOI-3629 b is a hot Jupiter ($M_{2}=0.26 \pm 0.02~\mathrm{M_J}$ and $R_{2}=0.74 \pm 0.02~\mathrm{R_J}$) orbiting its host star on a nearly circular orbit with a period of \(3.936551_{-0.000006}^{+0.000005}\) days. 

To determine if the low eccentricity of the orbits is consistent with tidal evolution, we estimate the timescales for circularization using the formalism of \cite{Jackson2008}. For the tidal quality factors, we assume each hot Jupiter is comparable to Jupiter and adopt a value of \(Q_{p}=10^5\) \citep[see][]{Goldreich1966,Lainey2009,Lainey2016}. We adopt a nominal value of \(Q_{\star}=10^7\) for early M dwarfs based on the modeling of \cite{Gallet2017}. Using the orbit parameters from joint modeling of each system and the stellar parameters, the timescale for circularization is $<0.1$ Gyrs for both systems, suggesting that these systems should be consistent with a circular orbit. If we adopt a larger value of $Q_p=10^7-10^9$ \citep[based on upper limits from][]{Bonomo2017} the circularization timescale can exceed $10$ Gyr and these systems may be able to retain a non-zero eccentricity. The RVs are able to place upper $3\sigma$ limits on eccentricity are $e<0.12$ for TOI-3714 b and $e<0.20$ for TOI-3629 b, revealing that even if these systems were not fully circularized, the planets are on low-eccentricity orbits.

\begin{deluxetable*}{llccccc}
{\tabletypesize{\tiny}
\tablecaption{System Parameters for TOI-3714 \label{tab:3714par}}
\tablehead{\colhead{~~~Parameter} &
\colhead{Units} &
\colhead{Prior} &
\multicolumn{4}{c}{Value}
}
\startdata
\noalign{\vskip 1.5ex} Photometric Parameters & & & TESS & RBO (08-16) & RBO (11-19) & ARCTIC\\ \noalign{\vskip .8ex}
~~~Linear Limb-darkening Coefficient$^a$ & $q_1$ & $\mathcal{U}(0,1)$ & $0.4_{-0.1}^{+0.2}$ & $0.5 \pm 0.2$ & $0.5 \pm 0.2$ & $0.4 \pm 0.1$ \\
~~~Quadratic Limb-darkening Coefficient$^a$ & $q_2$ & $\mathcal{U}(0,1)$ & $0.4 \pm 0.2$ & $0.2_{-0.1}^{+0.2}$ & $0.2_{-0.1}^{+0.2}$ & $0.5 \pm 0.1$\\
~~~Photometric Jitter & $\sigma_{phot}$ (ppm) & $\mathcal{J}(10^{-6},10^{3})$ & $0.007_{-0.007}^{+1.765}$ & $0.003_{-0.003}^{+1.431}$ & $0.03_{-0.03}^{+11.3}$ & $20_{-20}^{+778}$ \\
~~~Dilution Factor & $D$ & $\mathcal{U}(0,2)$ & $0.87 \pm 0.02$ & $\cdots$ & $\cdots$ & $\cdots$ \\
\hline
\noalign{\vskip 1.5ex} RV Parameters & & & \multicolumn{2}{c}{HPF} & \multicolumn{2}{c}{NEID}\\ \noalign{\vskip .8ex}
~~~Systemic velocity & $\gamma~\mathrm{(m~s^{-1})}$ & $\mathcal{U}(-10^3,10^3)$ &
\multicolumn{2}{c}{$-3 \pm 7$} & \multicolumn{2}{c}{$-81 \pm 6$} \\
~~~RV Jitter & $\sigma_{RV}~\mathrm{(m~s^{-1})}$ & $\mathcal{J}(10^{-3},10^3)$ & \multicolumn{2}{c}{$4_{-4}^{+20}$} & \multicolumn{2}{c}{$7_{-3}^{+6}$}\\
\hline
\sidehead{Orbital Parameters:}
~~~Orbital Period & $P$ (days)  & $\mathcal{N}(2.15,0.01)$ & \multicolumn{4}{c}{$2.154849 \pm 0.000001$}\\
~~~Time of Transit Center & $T_C$ (BJD\textsubscript{TDB}) & $\mathcal{N}(2458840.51,0.01)$ & \multicolumn{4}{c}{$2458840.5093 \pm 0.0004$}\\
~~~$\sqrt{e}\cos\omega$ &  & $\mathcal{U}(-1,1)$ & \multicolumn{4}{c}{$0.0 \pm 0.1$}\\
~~~$\sqrt{e}\sin\omega$ &  & $\mathcal{U}(-1,1)$ & \multicolumn{4}{c}{$0.1 \pm 0.1$}\\
~~~Semi-amplitude velocity & $K~\mathrm{(m~s^{-1})}$  & $\mathcal{J}(1,10^3)$ &   \multicolumn{4}{c}{$169_{-5}^{+6}$}\\
~~~Scaled Radius & $R_{p}/R_{\star}$  & $\mathcal{U}(0,1)$ & \multicolumn{4}{c}{$0.204 \pm 0.003$}\\
~~~Impact Parameter & $b$ & $\mathcal{U}(0,1)$ & \multicolumn{4}{c}{$0.26_{-0.1}^{+0.08}$}\\
~~~Scaled Semi-major Axis & $a/R_{\star}$  & $\mathcal{J}(1,100)$ & \multicolumn{4}{c}{$11.5_{-0.5}^{+0.4}$}\\
\hline
\sidehead{Gaussian Process Hyperparameters:}
~~~$B$ & Amplitude (ppm) & $\mathcal{J}(10^{-4},10^{12})$ & \multicolumn{4}{c}{$1.8_{-0.7}^{+2.3}$}\\
~~~$C$ & Additive Factor  & $\mathcal{J}(10^{-3},10^3)$ & \multicolumn{4}{c}{$1_{-1}^{+50}$}\\
~~~$L$ & Length scale (days)  & $\mathcal{J}(10^{-3},10^3)$ & \multicolumn{4}{c}{$3_{-2}^{+6}$}\\
~~~$P_{GP}$ & Period (days)  & $\mathcal{J}(1.0,100)$ & \multicolumn{4}{c}{$11_{-8}^{+30}$}\\
\hline
\sidehead{Derived Parameters:}
~~~Eccentricity & $e$  & $\cdots$ & \multicolumn{4}{c}{$0.03_{-0.02}^{+0.03}$, $3\sigma<0.12$}\\
~~~Argument of Periastron & $\omega$ (degrees)  & $\cdots$& \multicolumn{4}{c}{$100_{-200}^{+61}$}\\
~~~Orbital Inclination & $i$ (degrees) & $\cdots$ & \multicolumn{4}{c}{$88.7 \pm 0.5$}\\
~~~Transit Duration & $T_{14}$ (hours) & $\cdots$ & \multicolumn{4}{c}{$1.66_{-0.01}^{+0.02}$}\\
~~~Mass & $M_{p}$  ($\mathrm{M_{J}}$)  & $\cdots$ &  \multicolumn{4}{c}{$0.70 \pm 0.03$}\\
~~~Radius & $R_{p}$  ($\mathrm{R_{J}}$)  & $\cdots$ &  \multicolumn{4}{c}{$1.01 \pm 0.03$}\\
~~~Surface Gravity & $\log g_{p}$  (cgs)  & $\cdots$ &  \multicolumn{4}{c}{$3.25 \pm 0.04$}\\
~~~Density & $\rho_{p}$  ($\mathrm{g~cm}^{-3}$)  & $\cdots$ &  \multicolumn{4}{c}{$0.85 \pm 0.08$}\\
~~~Semi-major Axis & $a$ (au)  & $\cdots$ & \multicolumn{4}{c}{$0.027 \pm 0.001$}\\
~~~Average Incident Flux & $\langle F \rangle$ ($\mathrm{10^8\ erg~s^{-1}~cm^{-2}}$) & $\cdots$ & \multicolumn{4}{c}{$0.74_{-0.07}^{+0.06}$}\\
~~~Equilibrium Temperature\(^{b}\) & $T_{eq}$ (K) & $\cdots$ & \multicolumn{4}{c}{$750 \pm 20$}\\
\enddata
\tablenotetext{a}{Using the $q1$ and $q2$ parameterization from \cite{Kipping2013b}.}
\tablenotetext{b}{The planet is assumed to be a black body and we ignore heat redistribution.}
}
\end{deluxetable*}

\begin{deluxetable*}{llccccc}
{\tabletypesize{\tiny }
\tablecaption{System Parameters for TOI-3629 \label{tab:3629par}}
\tablehead{\colhead{~~~Parameter} &
\colhead{Units} &
\colhead{Prior} &
\multicolumn{4}{c}{Value}
}
\startdata
\noalign{\vskip 1.5ex} Photometric Parameters & & & TESS & RBO (09-26) & RBO (10-04) & Kuiper\\ \noalign{\vskip .8ex}
~~~Linear Limb-darkening Coefficient$^a$ & $q_1$ & $\mathcal{U}(0,1)$ & $0.4 \pm 0.2$ & $0.3_{-0.1}^{+0.2}$ & $0.6_{-0.3}^{+0.2}$ & $0.7 \pm 0.2$ \\
~~~Quadratic Limb-darkening Coefficient$^a$ & $q_2$ & $\mathcal{U}(0,1)$ & $0.3_{-0.2}^{+0.3}$ & $0.5 \pm 0.3$ & $0.3 \pm 0.2$ & $0.21 \pm 0.09$\\
~~~Photometric Jitter & $\sigma_{phot}$ (ppm) & $\mathcal{J}(10^{-6},10^{3})$ & $0.03_{-0.03}^{+423.61}$ & $0.001_{-0.001}^{+0.442}$ & $0.03_{-0.03}^{+16.45}$ &  $10_{-10}^{+421}$\\
~~~Dilution Factor & $D$ & $\mathcal{U}(0,2)$ & $0.90 \pm 0.04$ & $\cdots$ & $\cdots$ & $\cdots$ \\
\hline 
\noalign{\vskip 1.5ex} RV Parameters & & & \multicolumn{2}{c}{HPF} & \multicolumn{2}{c}{NEID}\\ \noalign{\vskip .8ex}
~~~Systemic velocity & $\gamma~\mathrm{(m~s^{-1})}$ & $\mathcal{U}(-10^3,10^3)$ & \multicolumn{2}{c}{$-15 \pm 3$} & \multicolumn{2}{c}{$-4_{-8}^{+10}$} \\
~~~RV Jitter & $\sigma_{RV}~\mathrm{(m~s^{-1})}$ & $\mathcal{J}(10^{-3},10^3)$ & \multicolumn{2}{c}{$5_{-3}^{+5}$} & \multicolumn{2}{c}{$16_{-7}^{+10}$}\\
\hline
\sidehead{Orbital Parameters:}
~~~Orbital Period & $P$ (days)  & $\mathcal{N}(3.94,0.01)$ & \multicolumn{4}{c}{$3.936551_{-0.000006}^{+0.000005}$}\\
~~~Time of Transit Center & $T_C$ (BJD\textsubscript{TDB}) & $\mathcal{N}(2458784.26,0.01)$ & \multicolumn{4}{c}{$2458784.256 \pm 0.001$}\\
~~~$\sqrt{e}\cos\omega$ &  & $\mathcal{U}(-1,1)$ & \multicolumn{4}{c}{$-0.1 \pm 0.1$}\\
~~~$\sqrt{e}\sin\omega$ &  & $\mathcal{U}(-1,1)$ & \multicolumn{4}{c}{$-0.1_{-0.1}^{+0.2}$}\\
~~~Semi-amplitude velocity & $K~\mathrm{(m~s^{-1})}$  & $\mathcal{J}(1,10^3)$ &   \multicolumn{4}{c}{$45 \pm 4$}\\
~~~Scaled Radius & $R_{p}/R_{\star}$  & $\mathcal{U}(0,1)$ & \multicolumn{4}{c}{$0.126 \pm 0.002$}\\
~~~Impact Parameter & $b$ & $\mathcal{U}(0,1)$ & \multicolumn{4}{c}{$0.2 \pm 0.1$}\\
~~~Scaled Semi-major Axis & $a/R_{\star}$  & $\mathcal{J}(1,100)$ & \multicolumn{4}{c}{$15.4 \pm 0.8$}\\
\hline
\sidehead{Gaussian Process Hyperparameters:}
~~~$B$ & Amplitude (ppm) & $\mathcal{J}(10^{-4},10^{12})$ & \multicolumn{4}{c}{$1.8_{-0.5}^{+1.5}$}\\
~~~$C$ & Additive Factor  & $\mathcal{J}(10^{-3},10^3)$ & \multicolumn{4}{c}{$3_{-3}^{+80}$}\\
~~~$L$ & Length scale (days)  & $\mathcal{J}(10^{-3},10^3)$ & \multicolumn{4}{c}{$0.8_{-0.4}^{+1.6}$}\\
~~~$P_{GP}$ & Period (days)  & $\mathcal{J}(1.0,100)$ & \multicolumn{4}{c}{$9_{-8}^{+37}$}\\
\hline
\sidehead{Derived Parameters:}
~~~Eccentricity & $e$  & $\cdots$ & \multicolumn{4}{c}{$0.05_{-0.04}^{+0.05}$, $3\sigma<0.20$}\\
~~~Argument of Periastron & $\omega$ (degrees)  & $\cdots$& \multicolumn{4}{c}{$-110_{-40}^{+200}$}\\
~~~Orbital Inclination & $i$ (degrees) & $\cdots$ & \multicolumn{4}{c}{$89.1 \pm 0.5$}\\
~~~Transit Duration & $T_{14}$ (hours) & $\cdots$ & \multicolumn{4}{c}{$2.20 \pm 0.03$}\\
~~~Mass & $M_{p}$  ($\mathrm{M_{J}}$)  & $\cdots$ &  \multicolumn{4}{c}{$0.26 \pm 0.02$}\\
~~~Radius & $R_{p}$  ($\mathrm{R_{J}}$)  & $\cdots$ &  \multicolumn{4}{c}{$0.74 \pm 0.02$}\\
~~~Surface Gravity & $\log g_{p}$  (cgs)  & $\cdots$ &  \multicolumn{4}{c}{$3.09_{-0.07}^{+0.06}$}\\
~~~Density & $\rho_{p}$  ($\mathrm{g~cm}^{-3}$)  & $\cdots$ &  \multicolumn{4}{c}{$0.8 \pm 0.1$}\\
~~~Semi-major Axis & $a$ (au)  & $\cdots$ & \multicolumn{4}{c}{$0.043 \pm 0.002$}\\
~~~Average Incident Flux & $\langle F \rangle$ ($\mathrm{10^8\ erg~s^{-1}~cm^{-2}}$) & $\cdots$ & \multicolumn{4}{c}{$0.53 \pm 0.06$}\\
~~~Equilibrium Temperature\(^{b}\) & $T_{eq}$ (K) & $\cdots$ & \multicolumn{4}{c}{$690 \pm 20$}\\
\enddata
\tablenotetext{a}{Using the $q1$ and $q2$ parameterization from \cite{Kipping2013b}.}
\tablenotetext{b}{The planet is assumed to be a black body and we ignore heat redistribution.}
}
\end{deluxetable*}

\section{Discussion}\label{sec:discussion}
\subsection{Constraints on unresolved stellar companions}\label{sec:secondarylight}
For both targets, the stellar density derived from the transit fit \citep[see][]{Seager2003,Winn2010} in Section \ref{sec:modelfit} is consistent with the value derived with an SED fit (Section \ref{sec:stellarpar}). Following the methodology presented in \cite{Kanodia2020}, we place limits on any spatially unresolved stellar companions to our targets by quantifying the lack of flux from a secondary stellar object in the HPF spectra. The highest S/N spectrum for each target is parameterized as a linear combination of a primary M dwarf\footnote{GJ\_205 and BD+29\_2279 for TOI-3714 and TOI-3629, respectively, as identified by \texttt{HPF-SpecMatch}} and a secondary stellar companion. The flux ratio between the secondary and primary star, $F$, is calculated as:
\begin{eqnarray}
S_{\mathrm{obs}} &=& A \left( (1-x)S_{\mathrm{primary}} + (x)S_{\mathrm{secondary}} \right) \label{eq:spectra} \\
F &=& \frac{x}{1-x} \label{eq:fluxratio}
\end{eqnarray}
\noindent where $S_{\mathrm{obs}}$ is the observed spectrum,  $S_{\mathrm{primary}}$ is the primary spectrum, $S_{\mathrm{secondary}}$ represents the secondary spectrum, and $A$ is the normalization constant. For a given primary and secondary template, we (i) shift the secondary spectrum in velocity space, (ii) add this shifted spectrum to the primary spectrum, and (iii) fit for the value of $x$ that best fits the observed spectrum. We limit the secondary spectral type to M dwarfs earlier than M7 and to the orders where telluric absorption is minimal. 

\begin{figure*}[!ht]
\fig{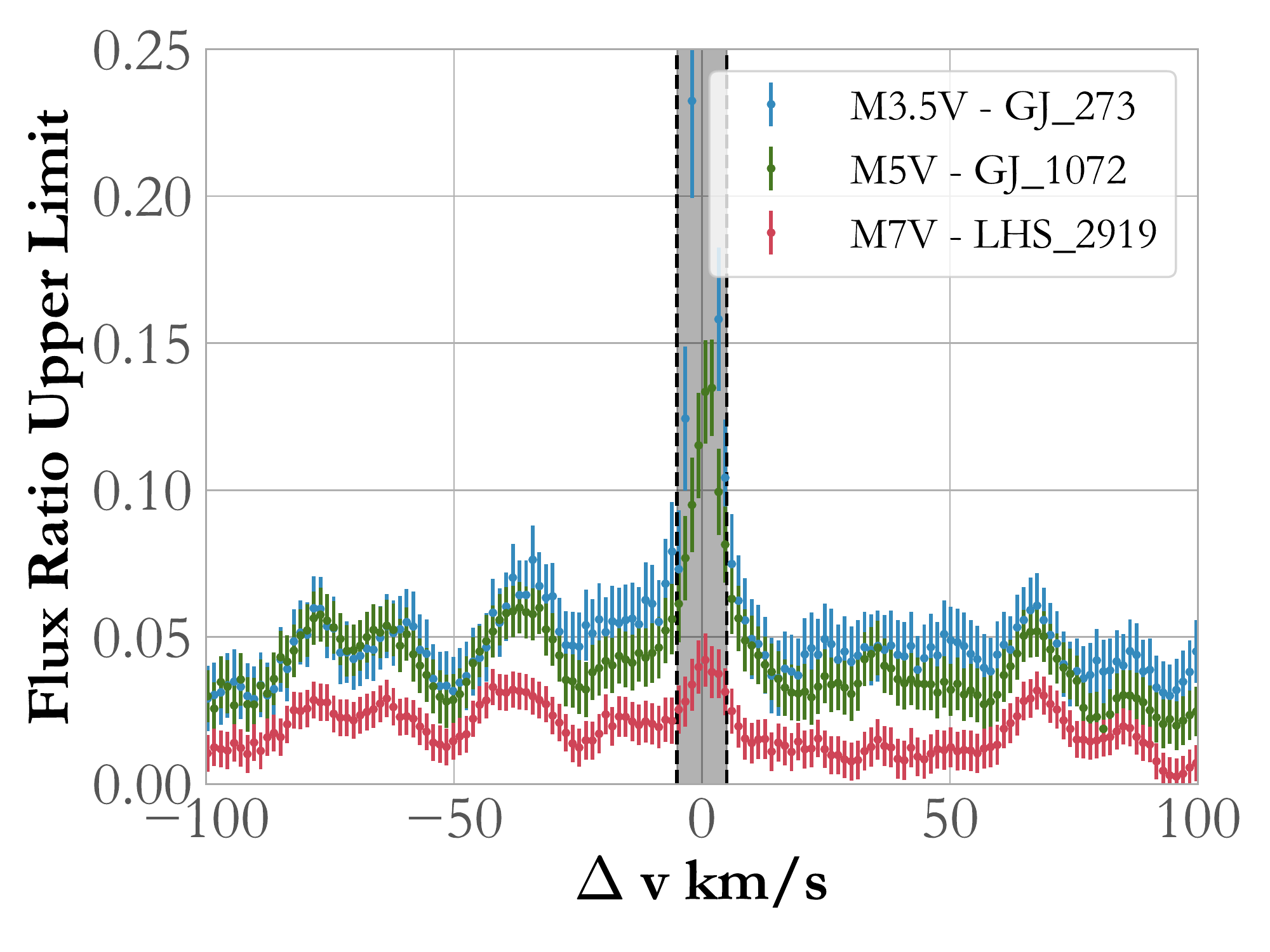}{0.45\textwidth}{{\small (a) Flux ratio upper limits for TOI-3714.}} 
\fig{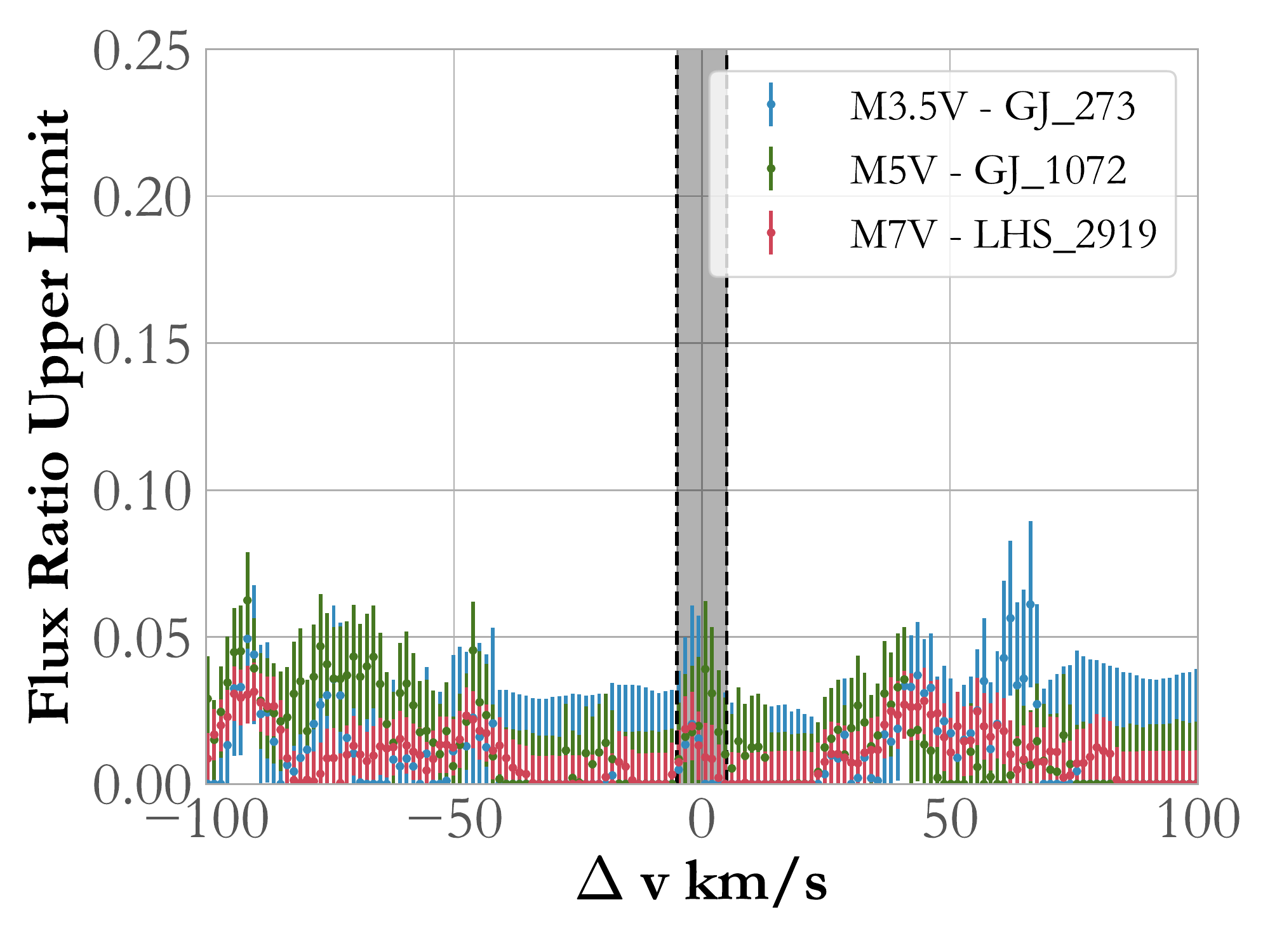}{0.45\textwidth}{ \small (b) Flux ratio upper limits for TOI-3629.}
\caption{\small  Flux upper limits placed on the flux ratio of a secondary companion to a template spectrum as a function of $\Delta \rm{v}$, obtained by fitting the wavelength region spanning $\sim 10450 - 10580$ \AA (HPF order index 17). We include the $1\sigma$ error bars, and shade the region corresponding to $\pm 5 ~\mathrm{km~s^{-1}}$. We place a conservative upper limit on the flux ratio of 0.07 for an unresolved stellar companion at separations $|\Delta v| > 5~\mathrm{km~s^{-1}}$.}\label{fig:secondarylight}
\end{figure*}

Figure \ref{fig:secondarylight} presents the results from HPF order index 17 spanning $10450 - 10580$ \AA. We place a conservative upper limit for a secondary of flux ratio = 0.07 or $\Delta \rm{mag} \simeq 2.9$ for both TOI-3714 and TOI-3629. As shown in Figure \ref{fig:secondarylight}, there is no significant flux contamination at $|\Delta v|$  $> 5~\mathrm{km~s^{-1}}$. We perform this secondary light analysis for velocity offsets from $5-100~\mathrm{km~s^{-1}}$, where the lower limit coincides with the spectral resolution of HPF ($\sim 5.5~ \rm{km~s^{-1}}$). The degeneracy between the primary and secondary spectra at velocity offsets $< 5\mathrm{~km~s^{-1}}$ prevent any meaningful flux ratio constraints at those velocity offsets.

\subsection{Constraints on unresolved bound companions}
We use {\tt thejoker} \citep{Price-Whelan2017} to perform a rejection sampling analysis on the residuals for the HPF RVs to constrain the existence of additional signals within the HPF RVs. This analysis used a log-uniform prior for the period (between 1 day and twice the HPF RV baseline), the Beta distribution from \cite{Kipping2013a} as a prior for the eccentricity, and a uniform prior for the argument of pericenter and the orbital phase. For both TOI-3714 and TOI-3629, we analyzed \(>10^8\) (\(2^{28}\)) samples with {\tt thejoker} and had a total acceptance rate of \(<3\%\). The surviving samples place an upper limit on any low-inclination ($\sin i\sim1$) companions of $M<3.1~\mathrm{M_J}$ ($K<\mathrm{300~m~s^{-1}}$) within 0.6 au ($P<242$ days) for TOI-3714 and $M<2.9~\mathrm{M_J}$ ($K<160\mathrm{~m~s^{-1}}$) within 1.4 au ($P<722$ days) for TOI-3629. 

Gaia EDR3 provides an additional constraint on the presence of close-in, massive companions with the re-normalized unit weight error (RUWE) statistic. \cite{Lindegren2021} note that the RUWE, or the square root of the reduced $\chi^2$ statistic that has been corrected for calibration errors, is sensitive to the photocentric motions of unresolved objects. In systems with massive companions on orbital periods much shorter than the baseline of Gaia (34 months for EDR3), the astrometric motion of the primary star around the center of mass may appear as noise when adopting a single-star astrometric solution \citep[e.g.,][]{Kervella2019,Kiefer2019}. \(\mathrm{RUWE}\gtrsim1.4\) is a threshold that correlates with the existence of an unresolved stellar companion in recent studies of stellar binaries \citep[e.g.,][]{Belokurov2020,Penoyre2020,Gandhi2020,Stassun2021}. With RUWE values of 1.15 and 1.05, Gaia EDR3 suggests TOI-3714 and TOI-3629 do not have massive stellar companions on short-periods ($\gtrsim0.1-3\mathrm{~years}$). Instead, these systems are in agreement with a single-star astrometric solution. 

\subsection{Constraints on resolved bound companions}
We also use results from Gaia EDR3 to determine if either star has a wide separation stellar companion. \cite{El-Badry2021} provide a list of spatially resolved binary stars from an analysis of proper motions. TOI-3629 is not contained in the catalog but TOI-3714 is identified as having a white dwarf stellar companion, Gaia EDR3 178924390476838784 (TIC 662037581). Systems in \cite{El-Badry2021} are flagged as having a white dwarf companion based on the location of the companion on the Gaia color-absolute magnitude diagram \citep{El-Badry2018b}. This object has a negligible probability ($\sim0.0006\%$) of being the chance alignment of a background source with spurious parallax and proper motion measurements. The white dwarf companion is located at a projected distance of 2.67\arcsec{} or a projected separation of 302 au from TOI-3714. This companion is outside both the HPF fiber \citep[$\sim1.7\arcsec$ on-sky;][]{Kanodia2018a} and the NEID HR fiber \citep[$\sim0.9\arcsec$ on-sky;][]{Schwab2016}.

\begin{figure*}[!ht]
    \epsscale{1.15} 
    \plotone{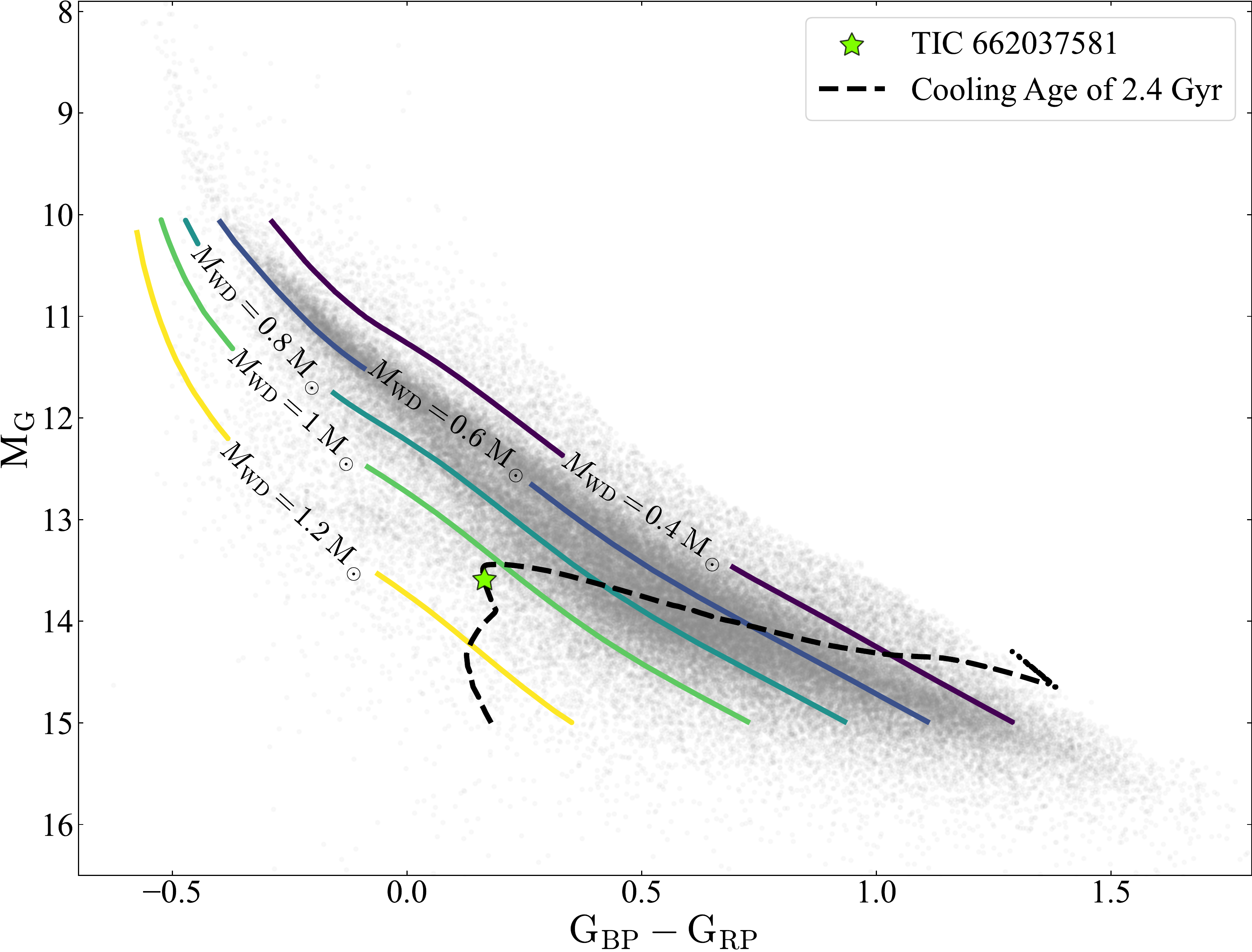}
    \caption{The nominal position of TIC 662037581, the white dwarf companion to TOI-3714, on the color-magnitude diagram for white dwarfs identified in Gaia EDR3 by \cite{GentileFusillo2021}. Contours for fixed masses from \cite{Bedard2020} are plotted for reference. The best-matching cooling track from the models is shown with a dashed line.} \label{fig:wdloc}
\end{figure*}    

To estimate the physical parameters of the white dwarf companion, we use the \texttt{WD\_models}\footnote{\url{https://github.com/SihaoCheng/WD_models}} package from \cite{Cheng2019} to derive a photometric age and mass from its location (see Figure \ref{fig:wdloc}) on the Gaia color-magnitude diagram \citep[data from][]{GentileFusillo2021}. We assume the atmosphere is composed of hydrogen and adopted the cooling models of \cite{Bedard2020}. We adopt these colors as nominal values but note that the proximity to TOI-3714 introduces some blending and contamination in the colors of the white dwarf companion, limiting the reliability of the estimated parameters. We use the \texttt{phot\_bp\_rp\_excess\_factor} as a diagnostic to determine if any of the measured blue or red Gaia photometry are problematic \citep{Evans2018, Riello2021}. We use Table 2 and Equation 6 from \cite{Riello2021} to calculate the corrected \texttt{phot\_bp\_rp\_excess\_factor}, which attempts to account for the color-dependent mean trend in this parameter. The corrected \texttt{phot\_bp\_rp\_excess\_factor} for the white dwarf companion is 0.81 and the deviation from zero suggests this object has some degree of contamination. 

Without additional photometry of TIC 662037581, we provide nominal parameters to qualitatively describe the companion. The estimated mass for the white dwarf companion is $M_{WD}\sim1.07~\mathrm{M_\odot}$ with a cooling age of $\sim2.4$ Gyr. The MIST semi-empirical white dwarf initial-final mass relationship from \cite{Cummings2018} suggests the progenitor star had a mass between $4.5-6.8~\mathrm{M_\odot}$. Stars in this mass range have typical lifetimes (pre-main sequence through post-asymptotic giant branch) of $<0.1$ Gyrs \citep{Dotter2016,Choi2016}. Assuming this object is coeval with TOI-3714, the combined progenitor lifetime and white dwarf cooling age ($\sim2.4$ Gyr) is consistent with the age range of $0.7-5.1$ Gyr estimated from the rotation period of TOI-3714 ($23.3 \pm 0.3$ days).

Approximately half of all hot Jupiter systems are known to have resolved stellar companions between separations of $50 - 2000$ au \citep[e.g.,][]{Wang2014a,Knutson2014a,Ngo2015,Ngo2016,Marzari2019,Hwang2020,Fontanive2021}, but only fourteen other exoplanetary systems \citep[see Table 2 in][]{Martin2021} are known to have a distant white dwarf companion. Of these fourteen systems, only the TOI-1259 \citep{Martin2021} and WASP-98 \citep{Hellier2014} systems host hot Jupiters. The existence of a distant stellar companion has been proposed as one mechanism to form hot Jupiters via a combination of secular interactions with the stellar companion and tidal friction \citep[e.g.,][]{Fabrycky2007,Anderson2016,Vick2019}. \cite{Ngo2016} note that most hot Jupiters with distant stellar companions are too separated to form via this mechanism. 

If the TOI-3714 system was initially a wide binary with an initial progenitor separation comparable to the observed separation ($\sim302$ au), the timescale for the Kozai cycles \citep[Equation 7 from][]{Kiseleva1998} would be \(\sim 2.8\) Gyr. This timescale is comparable to the age of the system and too long to effectively perturb a gas giant. The separation between the progenitor star and TOI-3714 could not have been too small, as a stellar binary with an initial separation $a\lesssim10$ au may interact when the primary star evolves off the main sequence and common envelope effects would subsequently shrink the orbit \citep[see][]{Paczynski1971,Paczynski1976,Ivanova2013}. Instead, the progenitor star could have been on a smaller orbit of tens of au and the onset of mass loss could have caused the orbit to expand \citep[see][]{Nordhaus2010,Nordhaus2013} to the observed separation. For example, at separations of 30 au, the Kozai timescale approaches $\sim3$ Myr and it may be possible the progenitor was close enough to perturb a nascent gas giant and far enough from TOI-3714 to prevent significant orbital decay. 

Gaia EDR3 is able to place constraints on the eccentricity of resolved wide binaries \citep[e.g.,][]{Tokovinin2020,Hwang2021}. The precision of Gaia EDR3 proper motion measurements allow for a measurement of the relative velocity for wide binaries (within the orbital plane) and allow a measurement of the angle between the separation vector and the relative
velocity vector (the $v-r$ angle). This is a function of the phase, inclination, eccentricity, and argument of pericenter \citep[see Appendix A in][for a detailed derivation]{Hwang2021}. The measured $v-r$ angle is $169\pm14^\circ$ and is significantly discrepant from a circular, face-on orbit ($v-r=90^\circ$). We follow the methodology and use the  software\footnote{\url{https://github.com/HC-Hwang/Eccentricity-of-wide-binaries}} described in \cite{Hwang2021} to estimate the posterior of the eccentricity distribution after adopting the parameters for the best-fitting power law \citep[Equation 29 in][]{Hwang2021} to the wide binary sample identified by \cite{El-Badry2021}. We note this eccentricity inference assumes that the wide companion has a random orbital orientation, an assumption which may not be true if the inner system is a transiting system. If the orbital orientation is not random, a large $v-r$ angle can indicate either a (i) high eccentricity or (ii) the outer companion lies on an orbit that is co-planar with the inner transiting system \citep[see Appendix B in][]{Hwang2020a,Behmard2022}. The inferred eccentricity with $1\sigma$ uncertainties for the orbit of the white dwarf companion is $e=0.99^{+0.01}_{-0.47}$. The high-eccentricity is consistent with the scenario in which the progenitor star was on a smaller orbit that widened and became eccentric due to mass loss. In this scenario, the resolved companion may have interacted with and impacted the migration of TOI-3714 b as it evolved into a white dwarf.

\subsection{Comparison to the M dwarf planet population}
\begin{figure*}[!ht]
\epsscale{1.15}
\plotone{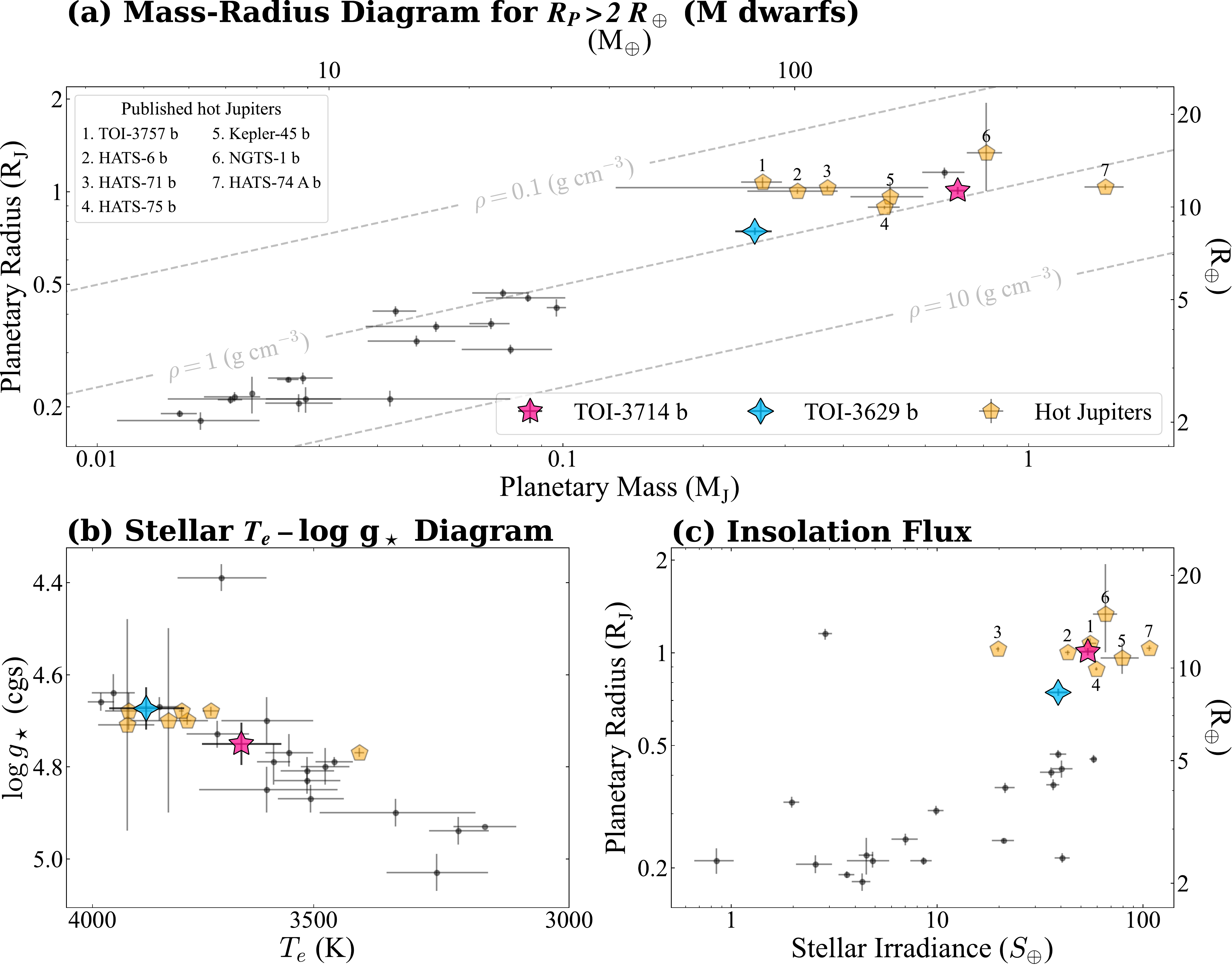}
\caption{The physical parameters of the TOI-3714 and TOI-3629 systems. \textbf{(a)} places TOI-3714 b (star) and TOI-3629 b (diamond star) on the mass-radius diagram for transiting M dwarf exoplanets with mass measurements and $R_p>2~R_\oplus$. All previously known hot Jupiters ($P<10$ days and $R_P\ge8~\mathrm{R_\oplus}$) transiting M dwarfs are marked as pentagons with number linked to the planet name. Contours of fixed bulk density are plotted for reference. \textbf{(b)} highlights the position of TOI-3714 and TOI-3629 on an effective temperature $-$ surface gravity diagram. \textbf{(c)} presents the insolation flux for M dwarf exoplanets. The data were compiled from the \href{https://exoplanetarchive.ipac.caltech.edu/cgi-bin/TblView/nph-tblView?app=ExoTbls&config=PSCompPars}{NASA Exoplanet Archive} \citep{Akeson2013} on 2022 May 4.} 
\label{fig:planetprops}
\end{figure*}

With the discovery of TOI-3714 b and TOI-3629 b, there are 9 M dwarf systems hosting transiting hot Jupiters ($P<10$ days and $R_p>8\mathrm{R_\oplus}$). Figure \ref{fig:planetprops} compares the planetary mass-radius, stellar $T_{e} - \log g_\star$, and the insolation flux of M dwarfs with transiting planets that have $R>2~R_\oplus$. Of the transiting hot Jupiters orbiting M dwarfs, TOI-3629 b has the smallest radius and mass ($\sim0.9\mathrm{~R_{Saturn}}$ and $\sim0.9\mathrm{~M_{Saturn}}$ ), and is the second coolest hot Jupiter with an insolation flux of $S=39\pm2~S_{\oplus}$. TOI-3714 has a radius comparable to the median value ($1\mathrm{~R_J}$) and an insolation flux ($S=54\pm5~S_{\oplus}$) comparable to the median value ($\sim60~S_\oplus$) of the population of hot Jupiters transiting M dwarfs. TOI-3714 is, however, the only known M dwarf with both a transiting hot Jupiter and a resolved wide companion. 

All M dwarfs hosting short-period ($P<10$ days) Jupiter-sized gas giants, including TOI-3714 b and TOI-3629 b, are early M dwarfs (M0-M3, $3400\mathrm{~K}<T_e<4000\mathrm{~K}$). This may simply be an observational bias or a result of a small population size, but in the framework of core accretion, factors such as protoplanetary disk mass may impact the formation of gas giants \citep[e.g.,][]{Mordasini2012,Hasegawa2013,Hasegawa2014,Adibekyan2019}. M dwarf protoplanetary disks have lower masses than the disks around Sun-like stars \citep[e.g.,][]{Andrews2013,Mohanty2013,Stamatellos2015,Ansdell2017}; disk masses for these stars are typically below a few Jupiter masses \citep{Ansdell2017,Manara2018}, such that the efficiency of gas giant formation is expected to increase when orbiting more massive M dwarfs because the materials that form gas giant cores are more abundant compared to the low-mass protoplanetary disks around later M dwarfs.

\begin{figure*}[!ht]
    \epsscale{1.15}
    \plotone{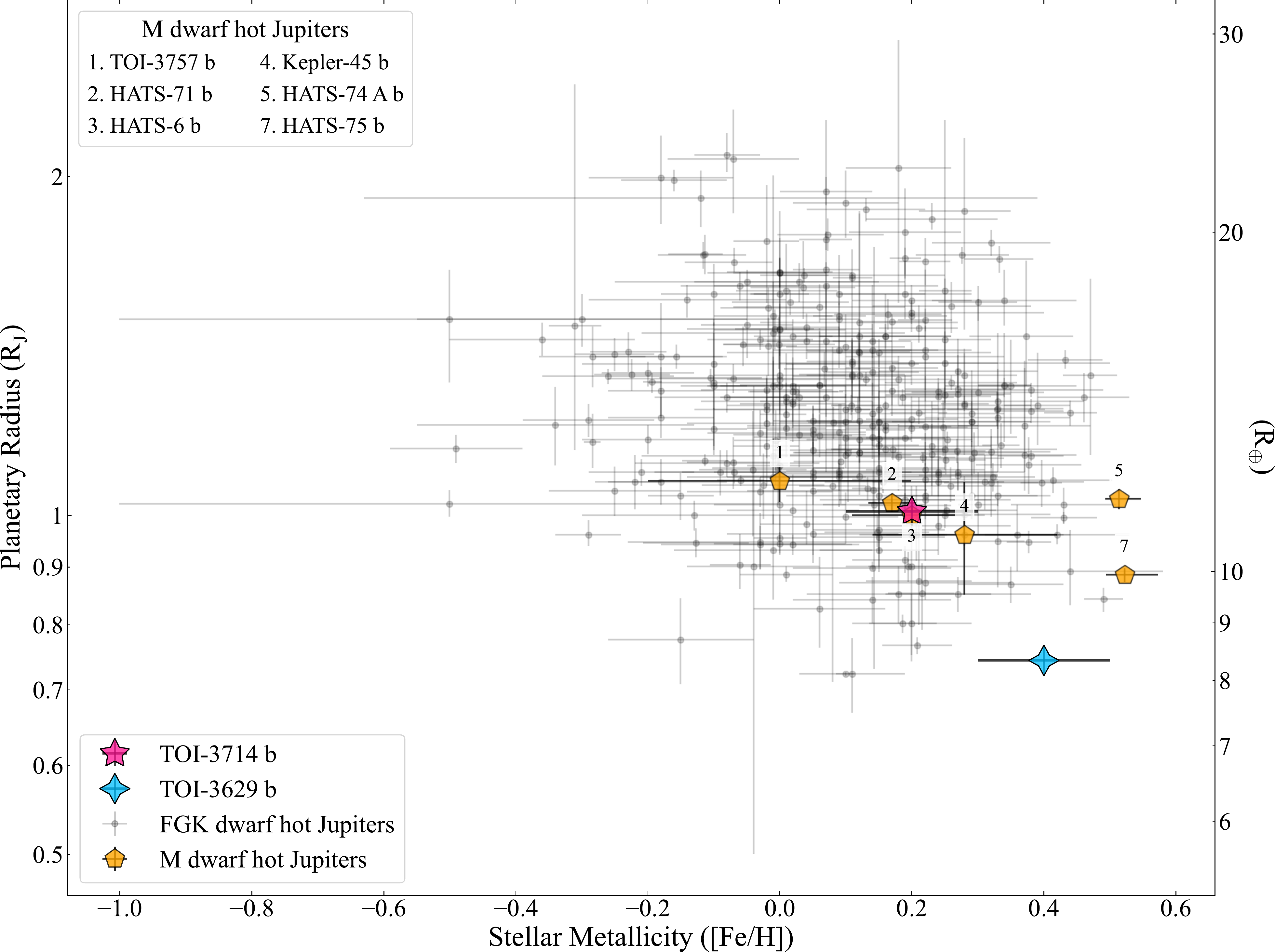}
    \caption{The position of TOI-3714 and TOI-3629 on a metallicity-mass diagram for transiting hot Jupiters ($R\ge8~\mathrm{R_\oplus}$ and $P<10$ days). All known hot Jupiters transiting M dwarfs have $\mathrm{[Fe/H]}>0$. The same numbers from Figure \ref{fig:planetprops} are included to identify the M dwarf hot Jupiters. HATS-6 b is behind the marker for TOI-3714 b. NGTS-1 b lacks a metallicity measurement \citep{Bayliss2018} and is not plotted. The data were compiled from the \href{https://exoplanetarchive.ipac.caltech.edu/cgi-bin/TblView/nph-tblView?app=ExoTbls&config=PSCompPars}{NASA Exoplanet Archive} on 2022 May 04.} 
    \label{fig:planetmetal}
\end{figure*}

In addition to the protoplanetary disk mass, the stellar metallicity has been known to be important for gas giant formation. The planet-metallicity correlation, in which metal-rich stars are more likely to host gas giant planets, has been extensively observed in Sun-like stars \citep[e.g.,][]{Fischer2005,Johnson2010,Guo2017,Osborn2020}. RV studies \citep[e.g.,][]{Neves2013,Maldonado2020} have suggested the planet-metallicity correlation exists for M dwarfs but there is no statistical study for M dwarfs with transiting gas giants because of the small population size. Figure \ref{fig:planetmetal} compares the metallicity and planetary radii of these two systems with transiting exoplanets from the NASA Exoplanet Archive. All M dwarfs hosting a transiting Jupiter-sized planet ($R\ge8~\mathrm{R_\oplus}$), including TOI-3714 and TOI-3629, have metallicities of $\mathrm{[Fe/H]}>0$. We perform a simple binomial probability calculation to assess how likely it is that all 10 Jupiter-sized companions are found to have $\mathrm{[Fe/H]}\ge0$ by random chance. If we assume a uniform distribution in metallicity between the range of $-0.5<\mathrm{[Fe/H]}<0.5$, the probability that all nine M dwarfs hosting hot Jupiters fall within the observed range of $\mathrm{[Fe/H]}\ge0$ is $\sim0.2\%$. 

We note that the metallicities on the NASA Exoplanet Archive are also not homogeneously derived. Metallicities derived using different techniques or instruments may exhibit offsets \citep[e.g.,][]{Guo2017,Petigura2018}. A magnitude limited search from a non-targeted transit survey, such as TESS, to identify a population of hot Jupiters orbiting M dwarfs is required to statistically evaluate if the planet-metallicity and stellar mass correlations apply to the population of transiting M dwarf gas giants. TESS is an ideal mission for this study, as it has been shown to be complete for hot Jupiters transiting earlier Sun-like stars \citep{Zhou2019a} and it should detect almost all transiting hot Jupiter - M dwarf systems given the large transit depth. This population of hot Jupiters could then be extensively studied to provide statistical constraints on the existence and strength of correlations of gas giant planets with metallicity or stellar mass.

TOI-3714 and TOI-3629, like the six existing M dwarf systems with transiting hot Jupiters, also lack additional transiting planets on nearby orbits. We search for additional transiting companions in the TESS data for each system by using the transit least squares algorithm \citep[\texttt{TLS};][]{Hippke2019} after subtracting the best-fitting transit model for each planet. For both TOI-3714 and TOI-3629, \texttt{TLS} only identifies candidate signals (depths $>1$ ppm) between $1-13$ days where the test statistic is below the suggested value of 7. This threshold corresponds to a false positive rate of $\sim1\%$ for the \texttt{TLS} algorithm. The maximum radius of a candidate signal identified by \texttt{TLS} was $\sim4\mathrm{~M_\oplus}$ and $\sim5\mathrm{~M_\oplus}$ for TOI-3714 and TOI-3629, respectively, such that the current TESS data excludes the existence of additional transiting gas giant companions in the TESS data.

Only five transiting hot Jupiters are known to exist in compact multiplanet systems: WASP-47 b \citep{Becker2015}, Kepler-730 b \citep{Zhu2018,Canas2019}, TOI-1130 c \citep{Huang2020a}, WASP-148 b \citep{Wang2022}, and WASP-132 b \citep{Hord2022}. This apparent low planetary multiplicity rate for hot Jupiters orbiting Sun-like stars has been detected in the analysis of multiple statistical samples from ground and space-based transiting hot Jupiters \citep[e.g.,][]{Steffen2012,Huang2016,Maciejewski2020,Hord2021,Wang2021a,Zhu2021}. The apparent lack of close-period companions to hot Jupiters may be imprints of high-eccentricity migration \citep[e.g.,][]{Mustill2015,Dawson2018}, as this mechanism would destabilize shorter period planets. TOI-3714 and TOI-3629 will both be observed in TESS cycle 5 and both sectors of data for each target could be analyzed in detail \citep[e.g., similar to][]{Hord2021} to provide robust constraints on additional transiting companions.

\subsection{Comparison to planetary models}
The equilibrium temperatures of TOI-3714 b ($T_{eq}=750\pm20$ K) and TOI-3629 b ($T_{eq}=690\pm20$ K) are $<1000$ K and it is unlikely these planets exhibit radius inflation due to stellar flux-driven mechanisms. Studies of the population of Kepler hot Jupiters \citep[e.g.,][]{Demory2011} determined that gas giants receiving an incident flux \(\lesssim 2 \times 10 ^{8}\mathrm{~erg~s^{-1}~cm^{-2}}\) have radii that are independent of the stellar incident flux. More recent analyses on transiting hot Jupiters \citep[e.g.,][]{Thorngren2018,Thorngren2021} have confirmed that inflated radii are evident in the population of hot Jupiters with $T_{eq}>1000$ K serving as a threshold for the onset of Ohmic heating and planetary inflation \citep[e.g.,][]{Batygin2010,Miller2011,Batygin2011}. 

Both the hot Jupiters TOI-3714 b and TOI-3629 b have $T_{eq}<1000$ K and do not show anomalously large radii when compared to models for gas giants from \cite{Baraffe2008} and \cite{Fortney2007}. The models from \cite{Fortney2007} assume a solar metallicity hydrogen and helium atmosphere with a heavy element core that is composed of a $50-50$ mixture of ice (water) and rock (olivine) while the models from \cite{Baraffe2008} assume a gaseous hydrogen and helium envelope with a distribution of heavy elements (water, dunite, and iron). Although these models are generated for the population of hot Jupiters transiting Sun-like stars, both TOI-3714 and TOI-3629 are in agreement with the non-irradiated models for gas giant interiors. We compared the observed radii of TOI-3714 b and TOI-3629 b to the radius predicted between $1-7$ Gyr models for a solar metallicity atmosphere and note agreement within $2-3\sigma$ regardless of age. The mass and radius of TOI-3714 b are in agreement with the models from \cite{Baraffe2008} containing a small fraction of heavy metals ($\sim2\%$) and with the \cite{Fortney2007} models for a core mass of $2-4\%$ of the planetary mass. The mass and radius of TOI-3629 are consistent with models from \cite{Fortney2007} having a core mass $\sim30\%$ of the planetary mass or the \cite{Baraffe2008} models with a heavy metal fraction of $\sim20-40\%$. These heavy metal fractions are consistent with what is seen in Jupiter \citep[$<10\%$ core mass;][]{Wahl2017} and Saturn \citep[$\sim20\%$;][]{Mankovich2021} in the Solar System. 

\subsection{Future Characterization}
\subsubsection{Stellar Obliquity}
The projected stellar obliquity ($\lambda$) is the apparent angle between the stellar rotation axis and the normal to the planet of the orbit. It can shed light on the dynamical and formation history of planets \citep[e.g.,][]{Albrecht2012,Winn2015,Triaud2018, Albrecht2021}. Measurements of $\lambda$ for hot Jupiters orbiting Sun-like stars \citep[e.g.,][]{Albrecht2012,Dawson2014} have revealed an obliquity distribution that is consistent with tidal realignment, indicating that their origin channels most likely involve dynamical interactions, such as planet-planet scattering. To date, there is no measurement of $\lambda$ for any M dwarf hosting a hot Jupiter. The measurement of $\lambda$ for either system via the Rossiter-McLaughlin (RM) effect \citep{Triaud2018} could limit the physical processes involved during formation because some mechanisms, such as disk migration, prohibit highly misaligned orbits \citep[see][]{Dawson2018}. 

The amplitude of the RM effect can be estimated as $\Delta V = 2/3 \left(R_{p}/R_\star\right)^2 v\sin i_\star \sqrt{1-b^2}$ \citep[Equation 1,][]{Triaud2018}. For TOI-3714, we estimate an equatorial ($\sin i=1$) rotational velocity of $v_{eq}=1.08 \pm 0.06~\mathrm{km~s^{-1}}$ using the derived rotation period and stellar radius. Additional photometric observations of TOI-3629 are required to determine the rotation period, however, if we adopt a value of $P_{rot}=30$ days corresponding to the marginally significant peak seen in its ZTF $zr$ data, the equatorial rotational velocity would be $v_{eq}= 1~\mathrm{km~s^{-1}}$. We use the derived transit parameters to estimate the RM effect amplitudes as \(\sim30~\mathrm{m~s^{-1}}\) and \(\sim10~\mathrm{m~s^{-1}}\) for TOI-3714 and TOI-3629, respectively. The precision to detect these amplitudes can be achieved using current high-resolution spectrographs with extended red wavelength coverage because both of these targets are early M dwarfs with a peak in the SED at around $\sim0.8-0.9$ microns.

\subsubsection{Transmission Spectroscopy}
TOI-3714 and TOI-3629 are the two brightest M dwarfs ($J<12$) with a transiting hot Jupiter and are potential targets to probe the atmosphere of warm ($T_{eq}\sim700$ K) M dwarf - hot Jupiter systems. \cite{Sing2016} obtained transmission spectra of hot Jupiters transiting Sun-like stars and noticed the observed sample contained both cloudy and clear planets, suggesting that hot Jupiters did not exhibit a strong relationship to cloud formation. While no extensive studies have been performed on M dwarf hot Jupiters, the transmission spectroscopy metric \citep[TSM;][]{Kempton2018} suggests that both TOI-3714 b ($\mathrm{TSM}=98\pm7$) and TOI-3629 b  ($\mathrm{TSM}=80\pm9$) are amenable to observations with the James Webb Space Telescope \citep[JWST;][]{Gardner2006}. These systems also have the precision on mass and radius (both determined at $>10\sigma$) needed for detailed atmospheric analysis \citep{Batalha2019}. It may be possible to determine atmospheric abundances of C-, N-, and O-bearing molecules in the atmospheres of these planets to probe the thermal structure of the interior \citep{Fortney2020}. While these hot Jupiters do not have the highest TSM of the existing population (TOI-3757 has the highest TSM of $180\pm30$), they are unique in the population as TOI-3629 b is the smallest hot Jupiter orbiting an M dwarf while TOI-3714 is one of the coolest M dwarfs hosting a hot Jupiter.   

TOI-3714 and TOI-3629 provide an opportunity to examine the prevalence of clouds and photochemical hazes for M dwarf exoplanets. Under certain combinations of temperature and surface gravity, clouds or hazes may form in the visible region of a hot Jupiter atmosphere either through condensation chemistry or photochemical processes \citep[e.g.,][]{Sudarsky2003,Helling2008,Marley2013} and the presence of clouds or hazes may weaken or mask spectral features \citep{Sing2016,Sing2018}. Photochemical processes are more efficient in cooler exoplanets \citep[e.g.,][]{Moses2011} and high incident stellar UV irradiation is thought to enhance the photochemical production of hydrocarbon aerosol \citep[e.g.,][]{Liang2004,Line2010}. Transmission spectra of TOI-3714 b and TOI-3629 b with JWST would probe atmospheric chemistry of gas giants orbiting M dwarfs and the effects of higher UV radiation environment of early M dwarfs on atmospheric chemistry \citep[e.g.,][]{Pineda2021}.

\section{Summary}\label{sec:summary}
We report the discovery of two gas giants orbiting M dwarfs. TOI-3714 b is a hot Jupiter ($M_{p}=0.70 \pm 0.03\mathrm{M_J}$ and $R_{p}=1.01 \pm 0.03\mathrm{R_J}$) on a $P=2.154849 \pm 0.000001$ day orbit. TOI-3629 b is a hot Jupiter ($M_{p}=0.26 \pm 0.02\mathrm{M_J}$ and $R_{p}=0.74 \pm 0.02\mathrm{R_J}$) on a $P=3.936551_{-0.000006}^{+0.000005}$ day orbit. Only TOI-3714 has a detectable rotation period of \(23.3 \pm 0.3\) days and most probably has an age between \(0.7-5.1\) Gyrs which is comparable to the nominal cooling age of its white dwarf companion ($\sim2.4$ Gyr). All hot Jupiters known to transit M dwarfs, including TOI-3714 and TOI-3629, orbit metal-rich early M dwarfs (M0-M3). A larger population size and homogeneously derived metallicities are required to confirm if the correlations with metallicity and stellar mass observed for hot Jupiters orbiting Sun-like stars are also observed in the population of M dwarf gas giants. Constraints from Gaia EDR3 and RVs reject the presence of massive short-period companions to both gas giants, but TOI-3714 has a resolved white dwarf companion at a projected separation of \(\sim300\) au and most likely on an eccentric orbit. The progenitor may have been close enough to impact the orbit of a nascent TOI-3714 b as it evolved into a white dwarf. TOI-3714 and TOI-3629 are the brightest M dwarfs hosting hot Jupiters ($J<12$) and are amenable to observations during transit to (i) further our understanding of their dynamical history with a measurement of the projected obliquity and (ii) explore the atmospheric chemistry of hot gas giants orbiting cool stars.

{\vskip6pt{\large\it Acknowledgments:}} We thank the anonymous referee for valuable feedback which has improved the quality of this manuscript. We thank Kareem El-Badry and David V. Martin for useful discussions. CIC acknowledges support by NASA Headquarters under the NASA Earth and Space Science Fellowship Program through grant 80NSSC18K1114, the Alfred P. Sloan Foundation's Minority Ph.D. Program through grant G-2016-20166039, and the Pennsylvania State University's Bunton-Waller program. 
The Center for Exoplanets and Habitable Worlds is supported by the Pennsylvania State University and the Eberly College of Science.
The computations for this research were performed on the Pennsylvania State University's Institute for Computational and Data Sciences' Roar supercomputer, including the CyberLAMP cluster supported by NSF grant MRI-1626251. This content is solely the responsibility of the authors and does not necessarily represent the views of the Institute for Computational and Data Sciences.
HCH acknowledges the support of the Infosys Membership at the Institute for Advanced Study. TNS acknowledges support from the Wyoming Research Scholars Program. 

The Pennsylvania State University campuses are located on the original homelands of the Erie, Haudenosaunee (Seneca, Cayuga, Onondaga, Oneida, Mohawk, and Tuscarora), Lenape (Delaware Nation, Delaware Tribe, Stockbridge-Munsee), Shawnee (Absentee, Eastern, and Oklahoma), Susquehannock, and Wahzhazhe (Osage) Nations.  As a land grant institution, we acknowledge and honor the traditional caretakers of these lands and strive to understand and model their responsible stewardship. We also acknowledge the longer history of these lands and our place in that history.

We acknowledge support from NSF grants AST 1006676, AST 1126413, AST 1310875, AST 1310885, AST 2009554, AST 2009889, AST 2108512 and the NASA Astrobiology Institute (NNA09DA76A) in our pursuit of precision RVs in the near-infrared. We acknowledge support from the Heising-Simons Foundation via grant 2017-0494.
We acknowledge support from NSF grants AST 1907622, AST 1909506, AST 1909682, AST 1910954 and the Research Corporation in connection with precision diffuser-assisted photometry.

This work is Contribution 0046 from the Center for Planetary Systems Habitability at the University of Texas at Austin.  These results are based on observations obtained with HPF on the HET. The HET is a joint project of the University of Texas at Austin, the Pennsylvania State University, Ludwig-Maximilians-Universit\"at M\"unchen, and Georg-August Universit\"at Gottingen. The HET is named in honor of its principal benefactors, William P. Hobby and Robert E. Eberly. The HET collaboration acknowledges the support and resources from the Texas Advanced Computing Center. We are grateful to the HET Resident Astronomers and Telescope Operators for their valuable assistance in gathering our HPF data.
We would like to acknowledge that the HET is built on Indigenous land. Moreover, we would like to acknowledge and pay our respects to the Carrizo \& Comecrudo, Coahuiltecan, Caddo, Tonkawa, Comanche, Lipan Apache, Alabama-Coushatta, Kickapoo, Tigua Pueblo, and all the American Indian and Indigenous Peoples and communities who have been or have become a part of these lands and territories in Texas, here on Turtle Island.

Some of the data presented were obtained by the NEID spectrograph built by the Pennsylvania State University and operated at the WIYN Observatory by NOIRLab, which is managed by the Association of Universities for Research in Astronomy (AURA) under a cooperative agreement with the NSF, and operated under the NN-EXPLORE partnership of NASA and the NSF. Observations with NEID were obtained under proposals 2021B-0035 (PI: S. Kanodia), 2021B-0435 (PI: S. Kanodia), and 2021B-0438 (PI: C. Ca\~nas). NEID results included here utilize the Data Reduction Pipeline operated by NExScI and developed under subcontract 1644767 between JPL and the University of Arizona. This work was performed for the Jet Propulsion Laboratory, California Institute of Technology, sponsored by the United States Government under the Prime Contract 80NM0018D0004 between Caltech and NASA.
WIYN is a joint facility of the University of Wisconsin-Madison, Indiana University, NSF's NOIRLab, the Pennsylvania State University, Purdue University, University of California-Irvine, and the University of Missouri. 
The authors are honored to be permitted to conduct astronomical research on Iolkam Du'ag (Kitt Peak), a mountain with particular significance to the Tohono O'odham. 

Some of results are based on observations obtained with the Apache Point Observatory 3.5m telescope, which is owned and operated by the Astrophysical Research Consortium. We wish to thank the APO 3.5m telescope operators in their assistance in obtaining these data.

Some of the observations in this paper made use of the NN-EXPLORE Exoplanet and Stellar Speckle Imager (NESSI). NESSI was funded by the NASA Exoplanet Exploration Program and the NASA Ames Research Center. NESSI was built at the Ames Research Center by Steve B. Howell, Nic Scott, Elliott P. Horch, and Emmett Quigley.

Some of the data presented in this paper were obtained from MAST at STScI. Support for MAST for non-HST data is provided by the NASA Office of Space Science via grant NNX09AF08G and by other grants and contracts.
This work includes data collected by the TESS mission, which are publicly available from MAST. Funding for the TESS mission is provided by the NASA Science Mission directorate. 
This research made use of the (i) NASA Exoplanet Archive, which is operated by Caltech, under contract with NASA under the Exoplanet Exploration Program, (ii) SIMBAD database, operated at CDS, Strasbourg, France, (iii) NASA's Astrophysics Data System Bibliographic Services, (iv) NASA/IPAC Infrared Science Archive, which is funded by NASA and operated by the California Institute of Technology, and (v) data from 2MASS, a joint project of the University of Massachusetts and IPAC at Caltech, funded by NASA and the NSF.

This work has made use of data from the European Space Agency (ESA) mission Gaia (\url{https://www.cosmos.esa.int/gaia}), processed by the Gaia Data Processing and Analysis Consortium (DPAC, \url{https://www.cosmos.esa.int/web/gaia/dpac/consortium}). Funding for the DPAC has been provided by national institutions, in particular the institutions participating in the Gaia Multilateral Agreement.

Some of the observations in this paper made use of the Guoshoujing Telescope (LAMOST), a National Major Scientific Project built by the Chinese Academy of Sciences. Funding for the project has been provided by the National Development and Reform Commission. LAMOST is operated and managed by the National Astronomical Observatories, Chinese Academy of Sciences.

Some of the observations in this paper were obtained with the Samuel Oschin Telescope 48-inch and the 60-inch Telescope at the Palomar Observatory as part of the ZTF project. ZTF is supported by the NSF under Grant No. AST-2034437 and a collaboration including Caltech, IPAC, the Weizmann Institute for Science, the Oskar Klein Center at Stockholm University, the University of Maryland, Deutsches Elektronen-Synchrotron and Humboldt University, the TANGO Consortium of Taiwan, the University of Wisconsin at Milwaukee, Trinity College Dublin, Lawrence Livermore National Laboratories, and IN2P3, France. Operations are conducted by COO, IPAC, and UW.

\facilities{ARC (ARCTIC), Exoplanet Archive, Gaia, HET (HPF), IRSA, KPNO:2.1m (Robo-AO), LAMOST, MAST, PO:1.2m (ZTF), PO:1.5m (ZTF), SO:Kuiper, TESS, WIYN (NEID, NESSI)} 
\software{
\texttt{astroquery} \citep{Ginsburg2019},
\texttt{astropy} \citep{AstropyCollaboration2018},
\texttt{barycorrpy} \citep{Kanodia2018}, 
\texttt{dynesty} \citep{Speagle2020},
\texttt{EXOFASTv2} \citep{Eastman2019},
\texttt{GLS} \citep{Zechmeister2009},
\texttt{HPF-SpecMatch} (S. Jones et al. 2022),
\texttt{juliet} \citep{Espinoza2019},
\texttt{lightkurve} \citep{LightkurveCollaboration2018},
\texttt{matplotlib} \citep{hunter2007},
\texttt{numpy} \citep{vanderwalt2011},
\texttt{pandas} \citep{McKinney2010},
\texttt{scipy} \citep{Virtanen2020},
\texttt{telfit} \citep{Gullikson2014},
\texttt{thejoker} \citep{Price-Whelan2017},
\texttt{TLS} \citep{Hippke2019},
\texttt{WD\_models} \citep{Cheng2019}
}

\bibliography{combined}
\bibliographystyle{aasjournal}



\end{CJK*}
\end{document}